\documentclass[physrev,reprint]{revtex4-2} %2-column mode

\usepackage{graphicx}% Include figure files
\usepackage{gensymb} %For \degree symbol
\usepackage{amssymb} % For square symbol
\usepackage[hidelinks]{hyperref} %For common footnotes
\usepackage{makecell} %To force line breaks in table cells and to change table style
\usepackage{textcomp} %for non-italic \mu
\usepackage{miller} %For crystallographic planes and directions
\usepackage{lineno}%For adding line numbers
\usepackage{xr}%For cross-referencing with different documents
\graphicspath{ {Images/} } %set path to graphic elements

\begin{document}
	% Use the \preprint command to place your local institutional report number 
	% on the title page in preprint mode.
	% Multiple \preprint commands are allowed.
	%\preprint{}
	
	\title[]{Surfactant behavior and limited incorporation of Indium during in-situ doping of GeSn grown by MBE}
		%Limitations of in-situ In doping of GeSn grown by MBE
	
	% Force line breaks with \\
	\author{A. Giunto}
	\author{L. E. Webb}
	\author{T. Hagger}
	\author{A. Fontcuberta i Morral}
	\email{anna.fontcuberta-morral@epfl.ch}
	\affiliation{ 
		Laboratory of Semiconductor Materials, Institute of Materials, École Polytechnique Fédérale de Lausanne, Lausanne, Switzerland
		%\\This line break forced with \textbackslash\textbackslash
	}%
	
	% repeat the \author .. \affiliation  etc. as needed
	% \email, \thanks, \homepage, \altaffiliation all apply to the current author.
	% Explanatory text should go in the []'s, 
	% actual e-mail address or url should go in the {}'s for \email and \homepage.
	% Please use the appropriate macro for the type of information
	
	% \affiliation command applies to all authors since the last \affiliation command. 
	% The \affiliation command should follow the other information.
	
	%\author{}
	%\email[]{Your e-mail address}
	%\homepage[]{Your web page}
	%\thanks{}
	%\altaffiliation{}
	%\affiliation{}
	
	% Collaboration name, if desired (requires use of superscriptaddress option in \documentclass). 
	% \noaffiliation is required (may also be used with the \author command).
	%\collaboration{}
	%\noaffiliation
	
	\date{\today}
	
	\begin{abstract}
		GeSn is a promising group-IV semiconductor material for on-chip Si photonics devices and high-mobility transistors.
		These devices require the use of doped GeSn regions, achieved preferably \emph{in-situ} during epitaxy.
		From the electronic valence point of view, p-type dopants of group-IV materials include B, Al, Ga and In. The latter element has never been investigated as p-type dopant in GeSn.
		In this work, we explore \emph{in-situ} In \emph{p-type} doping of GeSn grown by MBE.
		We demonstrate that In acts as a surfactant during epitaxial growth of GeSn:In, accumulating on surface and inducing segregation in the form of mobile Sn-In liquid droplets, strongly affecting the local composition of the material.
		In non-defective GeSn, we measure a maximal In incorporation of $2.8*10^{18}$cm$^{-3}$, which is two orders of magnitude lower than the values reported in the literature for \emph{in-situ p-type} doping of GeSn.
		We further show that In induces the nucleation of defects at low growth temperatures, hindering out-of-equilibrium growth processes for maximization of dopant incorporation.
		This work provides insights on the limitations associated with \emph{in-situ} In doping of GeSn, and discourages its utilization in GeSn-based optoelectronic devices.
	\end{abstract}
	
	%\pacs{68.35.bg, 68.37.Hk, 68.55.−a, 68.55.Ln, 68.55.A−}% insert suggested PACS numbers in braces on next line
	
	\maketitle 
	\section{\label{sec:introduction}Introduction}
	In the last decade, GeSn has been extensively studied as a novel semiconductor material for optoelectronic devices directly integrated on Si platforms~\cite{Moutanabbir2021,Miao2021,Zheng2018,Lin2017,Wirths2016}.
	Epitaxial integration of Ge$_{1-x}$Sn$_x$ on Si is enabled by their similar-sized diamond-like crystal structure, with a lattice mismatch equal to 4.2\% for $x=0$.
	The increase in lattice mismatch with increasing $x$ poses however challenges in the management and limitation of detrimental relaxation defects in the material.
	%However, the lattice mismatch increases for increasing $x$, inducing more relaxation defects, and practically limiting GeSn-on-Si epitaxy to alloy compositions with $x<0.3$, even with the use of graded buffers\cite{Zheng2018a}. 
	In addition, Ge$_{1-x}$Sn$_x$ is a metastable material for $x>0.01$, and thus its synthesis requires low-temperature, out-of-equilibrium processes where Ge-Sn phase separation is kinetically hindered.
	Above the material critical stability temperature, which is inversely related to the Sn fraction in the alloy, Sn segregates out of GeSn lattice, clustering in the bulk and forming mobile Sn droplets on the film surface~\cite{Mukherjee2021,Groiss2017,Zaumseil2018,Kuchuk2022}.
	
	In spite of these challenges, the promising optoelectronic properties of GeSn drove research efforts in the development of this material.
	Ge$_{1-x}$Sn$_x$ possesses a direct bandgap in the near- and short-wave infrared wavelengths  for approximately $x>8.5$\%at~\cite{Gallagher2014,Wirths2015}, motivating the fabrication of GeSn Light Emitting Devices (LEDs)~\cite{Cardoux2022,Huang2019,Chang2017}, lasers~\cite{Wirths2015,Kim2022,Ojo2022}, and photodiodes~\cite{Atalla2022,TalamasSimola2021,Tran2019} for on-chip Si photonics~\cite{Moutanabbir2021}.
	Furthermore, GeSn theoretical carrier mobility values larger than Si and Ge~\cite{Sau2007,Mukhopadhyay2021} pushed for the realization of GeSn high-mobility field-effect transistors (FET)~\cite{Liu2020,Huang2017,Wang2018}.
	These (opto)electronic devices necessitate \emph{p-type} and/or \emph{n-type} doped regions, obtained through incorporation in the GeSn lattice of elements from respectively columns III (B, Al, Ga, In) and V (P, As, Sb) of the periodic table.
	Incorporation of doping elements can be achieved through ion implantation, or \emph{in-situ} during growth of GeSn, whereas high-temperature diffusion processes to dope GeSn from the gas phase cannot be employed due to the material metastability.
	In general, to avoid lattice damage and amorphization from implantation, it is desirable to dope GeSn \emph{in-situ} during growth, though this method requires thorough investigation of growth parameters to accurately calibrate dopant concentrations in the film.
	In addition, one needs to verify that the introduced dopants are not detrimental for the film growth process.
	
	Several studies have demonstrated \emph{in-situ} doping of epitaxial GeSn by Chemical Vapor Deposition (CVD)~\cite{Vohra2019,Tsai2020,Frauenrath2021,Tsai2018,Margetis2017} and Molecular Beam Epitaxy (MBE)~\cite{Shimura2012,Taoka2017,Wang2016,Nakatsuka2013,Bhargava2014}.
	While both \emph{n-type} and \emph{p-type} doping have been realized, the focus of this paper is exclusively on \emph{p-type} doping.
	Previous works on \emph{in-situ p-type} doping of GeSn revolved around only two group-III dopant elements, namely B and Ga.
	The highest active \emph{p-type} dopant concentration of $3.2*10^{20}$~cm$^{-3}$ was reported both by Vohra \emph{et al.}~\cite{Vohra2019} for GeSn:B grown by CVD and by Wang \emph{et al.}~\cite{Wang2016} for GeSn:Ga grown by MBE.
	Though this maximal active concentration is in principle sufficient for most applications, it is desirable to explore different possibilities of \emph{in-situ p-type} doping, namely using In and Al as dopant elements.
	Their demonstration would increase the number of options available to fabricate (opto)electronic devices with different combinations of material systems (e.g. III-V/IV heterojunction tunnel FETs~\cite{Sathishkumar2020}).
	
	Furthermore, the addition of group-III elements to epitaxy of GeSn requires special attention, in that it may alter the growth dynamics with respect to the pure GeSn system.
	Elements from groups III and V are known to act as surfactants during growth of group-IV Ge and Si films~\cite{Voigtlander1995,Wang2004,Klatt1994}.
	By analogy, we could expect similar behavior for growth of GeSn, being itself from group IV.
	Surfactant elements tend to remain on surface during epitaxial growth: when buried by a surface monolayer, they exchange their position with the surface adatom; being this exchange process faster than the characteristic time for monolayer growth, surfactants manage to escape the subsurface layer before getting buried underneath newly grown monolayers~\cite{Pohl2020}.
	This signifies that surfactant doping elements tend not to incorporate in the growing film, and thus cannot dope it in significant concentrations.
	Nevertheless, despite the fact that the surfactant effect can strongly alter dopant incorporation, it is rarely discussed in the literature whether group III dopants act as surfactants during GeSn growth.
	To the extent of our knowledge, only Shimura \emph{et al.}~\cite{Shimura2012} took this phenomenon into consideration in their study on GeSn:Ga grown by MBE:
	While Ga acts as a surfactant in pure Ge epitaxial growth, they demonstrated the loss of its surfactant properties when Sn was added in the growth of GeSn. Hence, they could show that Ga is an optimal element for \emph{in-situ p-type} doping of GeSn.
	
	To expand the range of \emph{p-type} doping options for the GeSn system, in this study we explore \emph{in-situ} Indium \emph{p-type} doping of GeSn grown by MBE.
	Indium is known to possess a low solubility in pure Ge ($\sim4*10^{18}$cm$^{-3}$) compared to the other group-III dopant elements, e.g. Ga, with solubility of $4.9*10^{20}$cm$^{-3}$ \cite{Simoen2007}.
	Expecting a similar behavior in GeSn, In doping of the alloy would be in principle discouraged.
	However, this has never been demonstrated experimentally, and is thus worth investigating.
	In addition, in this work, we show that the In dopant element influences the growth dynamics of the GeSn alloy.
	We prove that the presence of In during growth enhances defect nucleation and facilitates the segregation of Sn, inducing the formation of In-Sn segregation droplets if the growth conditions are not properly selected.
	With secondary-ion mass spectroscopy (SIMS) characterization, we demonstrate that In acts as a surfactant in this growth system, accumulating on the GeSn surface as the film grows.
	Finally, we elucidate the possible thermodynamic contributions causing enhanced surface Sn segregation in presence of In, and we discuss the limitations of In doping of GeSn alloys.

	\section{\label{sec:experimental}Experimental Methods}
	Intrinsic Ge(001) substrates were exposed to a O$_2$ plasma to remove organic contaminants from surface, and were  dipped in HF 1\% for 90 seconds to remove the Ge native oxide. Substrates were then rinsed in DI water and dried with a N$_2$ blowing gun. After chemical cleaning, they were introduced in a \emph{Veeco GENxplor} MBE growth system, and degassed in the load-lock at 150\degree C for 30~minutes. The substrates underwent a further degassing step in a preparation module at 600\degree C before being introduced in the growth chamber. Here, they underwent a second deoxidation step at 750\degree C for 15 minutes.
	Epitaxial, monocrystalline GeSn:In films were grown by evaporating Ge, Sn and In from individual Knudsen cells.
	The base pressure of the growth chamber at the start of the growth was of the order of $1*10^{-10}$Torr.
	Substrate nominal temperatures were calibrated with an infrared thermal camera, and ranged from 185\degree C to 205\degree C, with uncertainty of $\pm20$\degree C.
	Growth parameters for the different samples grown in this work are reported in Tab.~\ref{tab_SampleDescriptions}, with additional details in Tab.~SI1.
	Growths of samples~E, F and I were repeated to confirm the reproducibility of the observed physical phenomena.
	
	Film morphology and thickness were characterized with a \emph{Zeiss Merlin} scanning electron microscope (SEM).
	Epitaxial relation of GeSn:In films with the Ge substrates was demonstrated by X-ray diffraction reciprocal space mapping (XRD RSM) with an \emph{X-ray Bruker D8 Discover}.
	The measured RSM, reported in Fig.~SI1, were used to calculate film composition and strain.
	Indium incorporation, being lower than 0.01\%at, does not appreciably influence the macroscopic lattice parameter of the GeSn alloy measured with XRD, and we thus used Vegard's law to calculate GeSn composition without any bowing correction \cite{Xu2017}.
	In a \emph{FEI Talos} system, monocrystallinity was confirmed by transmission electron microscopy (TEM), and local composition was probed with scanning TEM energy-dispersive x-ray spectroscopy (STEM~EDX). Both characterizations were performed on cross-sectional lamellae cut out of the film with a \emph{Zeiss NVision} focused ion beam (FIB).

	The In concentration of few samples was measured via SIMS depth profiling, performed by \emph{EAG laboratories}.
	To characterize the In concentration at the GeSn:In films surface, prior to SIMS analysis these samples were covered with 100nm of Ge deposited by electron-beam evaporation with no heating applied at the sample.
	The In concentration was calibrated using Ge:In standards as GeSn:In standards were unavailable, leading to a 15\% uncertainty on the reported In concentration values.
	SIMS profiling was also used to verify that the Sn composition across the film thickness was uniform (see Fig.~SI4), as expected from MBE-grown films with Sn concentrations below 10\%at \cite{Rathore2021}.

	\begin{table}
		\centering
		\caption{\label{tab_SampleDescriptions}MBE deposition parameters of Ge$_{1-x}$Sn$_x$:In films studied in this work. All films are monocrystalline, pseudomorphic on Ge(001).
		%Growth rates for samples C, D, E are respectively 19.7nm/min, 18.8nm/min and 18.0nm/min. The In flux is not expected to influence the growth rate.
		The Ge flux was fixed at 1000nTorr for all samples. Fluxes are reported in \emph{nTorr}, as per measurement from the MBE beam flux monitor.
		Minor effective flux variations from growth to growth resulted in slight variability in GeSn thickness and alloy compositions despite constant substrate temperature ($T$) and Sn/Ge flux ratios, e.g. samples~A, B and C.}
		\begin{tabular*}{0.48\textwidth}{c @{\extracolsep{\fill}}|c |c |c |c |c |c |c}
			\makecell{ID} &  \makecell{Sub. $T$\\ (\degree C)} &    \makecell{Sn flux \\ (nTorr)} &  \makecell{In flux \\ (nTorr)} &  \makecell{Gr. time \\ (min)}&   \makecell{Thick. \\ (nm)} & \makecell{$x_{Sn}$ \\ (\%at)} & \makecell{In-plane \\ str. (\%)} \\
			\hline
			A & 205  & 50 & 0.5 & 30 & 547 & 1.8 & -0.26 \\
			\hline
			B & 205  & 50 & 1 & 30 & 542 & 2.0 & -0.30 \\
			\hline
			C & 205  & 50 & 3 & 30 & 552 & 1.7 & -0.26 \\
			\hline
			D & 205  & 100 & 1 & 30 & 565 & 3.8& -0.56%\textsuperscript{\ref{foot_PartialSegregation}}  
			\\
			\hline
			E & 205  & 150 & 0.5 & 30 & 592 & 5.7 & -0.82 \\
			\hline
			F & 205  & 150 & 1 & 30 & 590 & 5.9 & -0.85%\footnote{\label{foot_PartialSegregation}Sn is partially segregated out of the film, reducing the incorporated Sn content with respect to non-segregated films with the same Ge-Sn flux ratio.}
			\\
			\hline
			G & 205  & 150 & 3 & 30 & 571 & -\footnote{\label{foot_FullSegregation}Sn is  fully segregated out of the film. See SEM image in Fig.~\ref{fig_Co-SegregationTable}, and XRD RSM in Fig.~SI1.} & -\textsuperscript{\ref{foot_FullSegregation}.}   \\
			\hline
			H & 195  & 150 & 1 & 30 & 580 & 5.3& -0.78%\textsuperscript{\ref{foot_PartialSegregation}}  
			\\
			\hline
			I & 185  & 150 & 1 & 30 & 583 & 5.4 & -0.80 \\
			\hline
			J & 205  & 150 & 1 & 10 & 183 & 5.7 & -0.85 \\
			\hline
			K & 205  & 150 & 1 & 20 & 404 & 5.4 & -0.80%\textsuperscript{\ref{foot_PartialSegregation}}  
			\\
		\end{tabular*}
	\end{table}
	
	\begin{figure*}
		\includegraphics[width=\textwidth]{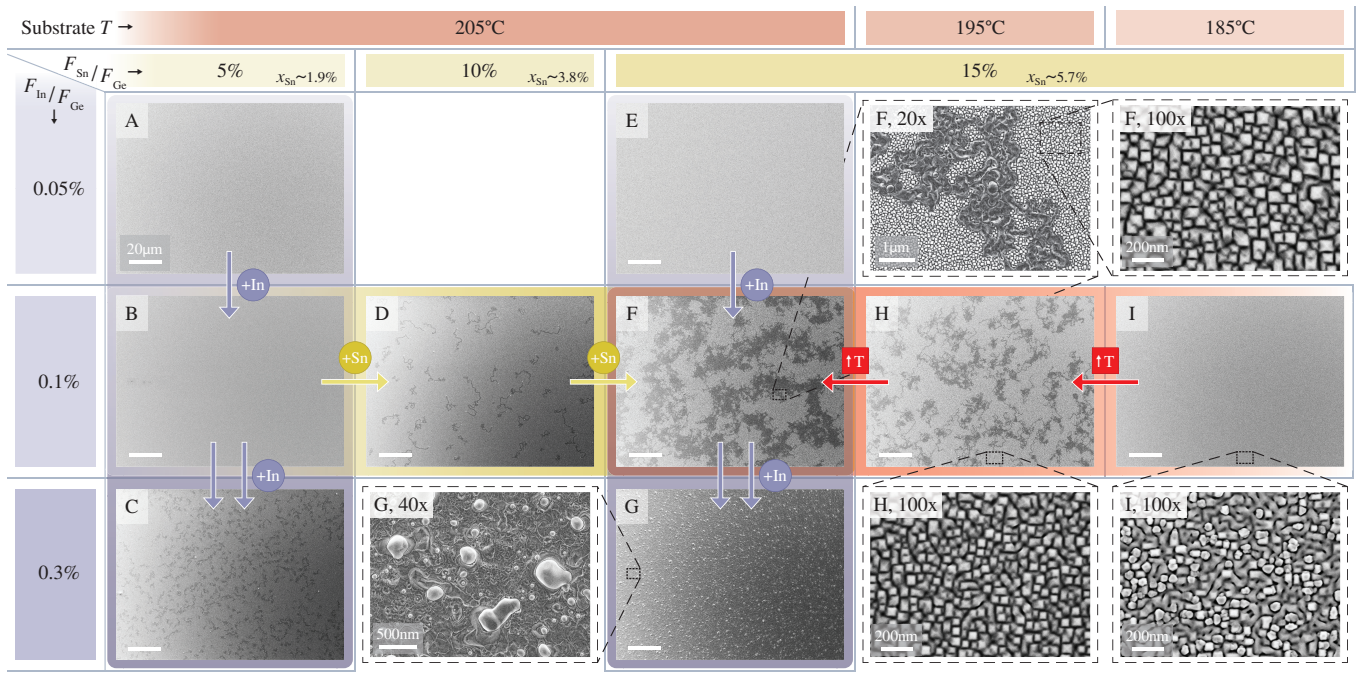}%Fig_Sn-In_fluxes.png}
		\centering
		\caption{\label{fig_Co-SegregationTable}SEM top-view images of monocrystalline GeSn:In films grown on Ge(001) substrates by MBE. Scale bars are 20\textmu m, unless differently specified.
		The first row shows the substrate growth temperature ($T$), while the first column and the second row show respectively the In/Ge and Sn/Ge flux ratios (respectively $F_{In}/F_{Ge}$, $F_{Sn}/F_{Ge}$) used during deposition of the different films.
		In the second header row we report also the approximate Sn atomic fraction ($x_{Sn}$) corresponding to the $F_{Sn}/F_{Ge}$.
		In the inset of each figure is the sample ID, as per Tab.~\ref{tab_SampleDescriptions}. A 20x magnification on sample~F shows segregation droplets and the characteristic trails they leave behind. These trails confer a darker SEM contrast that allows to evaluate the extent of segregation simply from top-view SEM imaging. Color schemes and arrows show that by increasing the substrate temperature, In flux, and/or Sn fraction, segregation increases, and eventually covers the entire sample surface, as in the case of sample~G.
		Images magnified by 100x of samples~F, H, I show the typical surface morphology of GeSn:In films at different substrate $T$.}
	\end{figure*}
	
	\section{\label{sec:results}Results}
	
	GeSn:In epitaxial films with uniform Sn composition were grown on Ge(001) substrates by MBE to investigate \emph{in-situ p-type} doping of GeSn by In.
	Growth parameters and film characteristics of these samples are reported in Tab.~\ref{tab_SampleDescriptions}.
	Except for sample~G, where Sn is almost fully segregated out of the film, all GeSn:In films are pseudomorphic, fully strained, the absolute value of in-plane compressive strain being determined by the GeSn alloy composition.
	Values of GeSn composition and in-plane strain for partially segregated samples C, D,  F, and H refer to the non-segregated regions of these samples.
	%The partial segregation in samples C, D, F, and J, seen in Sec.~\ref{sec:co-segregation}, does not affect the XRD RSM composition and strain measurements, which provide the characteristics of the non-segregated regions of the films.
	
	\subsection{\label{sec:co-segregation}Co-segregation of In and Sn}
	In Fig.~\ref{fig_Co-SegregationTable}, we report top-view SEM images of a combination of different epitaxial GeSn:In films on Ge(001), grown varying substrate temperature, GeSn alloy composition, and In dopant flux.
	The sample label at the top-left of each SEM image refers to the growth parameters reported in Tab.~\ref{tab_SampleDescriptions}. 
	Samples~C, D, F and H show regions of dark SEM contrast in a lighter background, while all other samples, with the exception of sample G, at low magnification present a surface with homogeneous light SEM contrast.
	The 20x magnified SEM image of sample F reveals the origin of dark SEM contrast: segregation droplets have formed and moved around during growth, leaving behind a trail that appears darker at SEM.
	This segregation behavior resembles closely that of pure GeSn, where Sn liquid droplets move on surface during growth, dissolving the GeSn film at their front and depositing behind a trail of almost pure Ge. This phenomenon has been accurately modeled in Refs.~\onlinecite{Groiss2017,Kuchuk2022}.
	
\begin{figure}
		\includegraphics[width=0.48\textwidth]{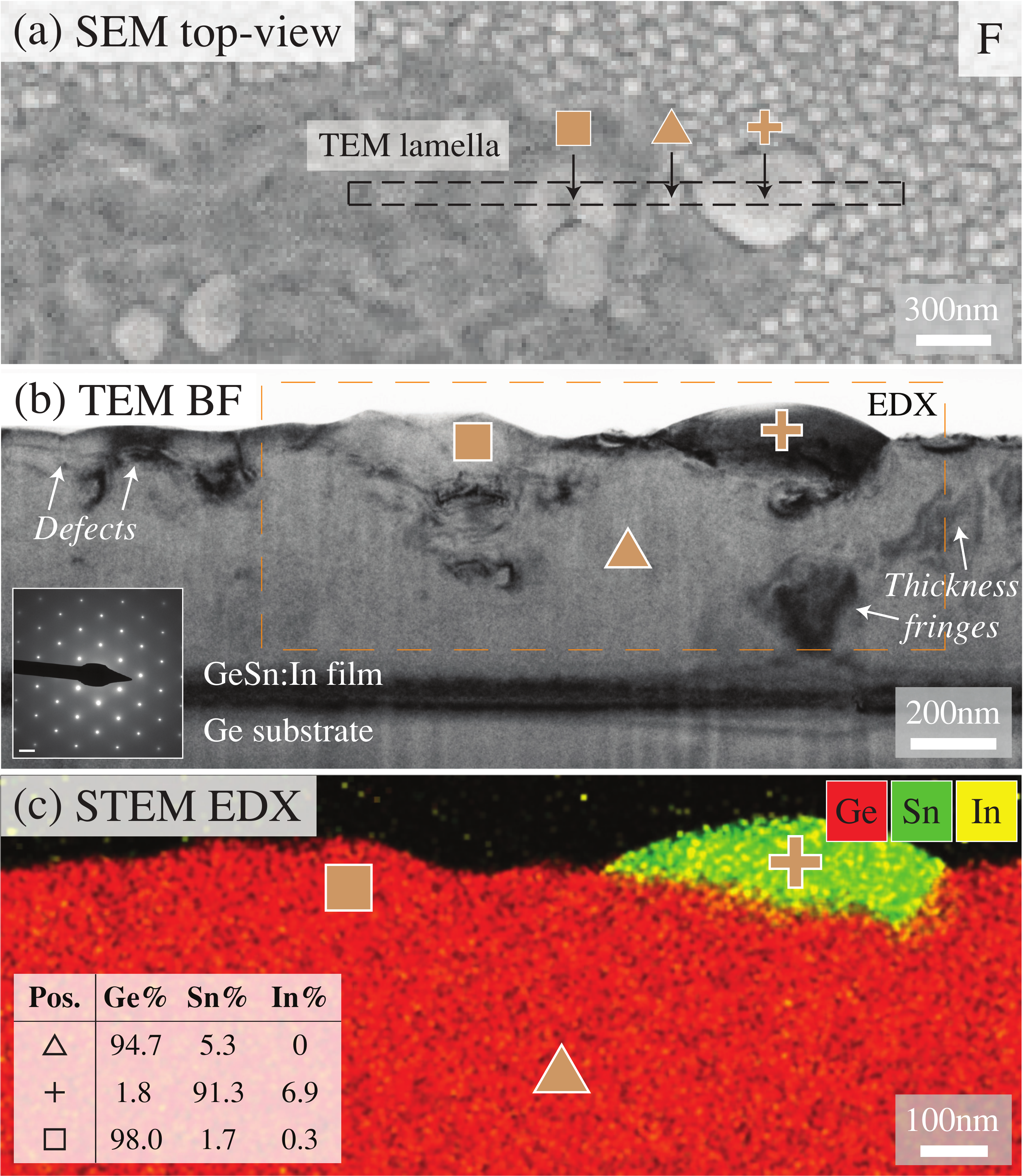}
		\caption{\label{fig_EDX} (a) Top-view SEM image of segregation droplets and trails on sample F.
		(b) TEM bright field image of the region framed withing a dashed, black rectangle in (a).
		The diffraction pattern in the inset, with scale bar of 2~nm$^{-1}$, demonstrates single crystallinity of the GeSn:In film.
		Orange symbols indicate the positions probed by STEM EDX in (c), with the corresponding measured atomic compositions reported in the table inset. 
		EDX shows that the segregation droplet (+) is mainly composed of Sn and In. The droplet trail ($\Box$) is instead  almost pure Ge, as expected from previous works. Full measurement details are reported in Fig.~SI8.}
	\end{figure}

	To understand the origin of segregation, we performed TEM characterization of a segregation region on sample~F, presented in Fig.~\ref{fig_EDX}.
	The top-view SEM image in Fig.~\ref{fig_EDX}a illustrates the surface morphology of sample F, more neatly visible in Fig.~\ref{fig_Co-SegregationTable}~(F,~20x and 100x) and in Fig.~SI7.
	A dashed, black rectangle in Fig.~\ref{fig_EDX}a indicates the region probed by TEM in Fig.~\ref{fig_EDX}b, which contains a segregation droplet, part of the droplet trail, and portions of non-segregated GeSn:In film.
	TEM bright-field (BF) imaging in Fig.~\ref{fig_EDX}b shows defects underneath the segregation regions (i.e., droplet and its trail), which seem to indicate a boundary with pristine GeSn:In film, confirmed to be monocrystalline in the inset TEM diffractogram.
	Segregation thus appears to occur only in the surface region, suggesting droplet nucleation takes place at a late stage of growth.
	In addition, we observe that the segregation droplet does not dissolve the entire film underneath, which remains unaffected. The TEM contrast visible underneath the droplet is in fact only due to thickness fringes (see also Fig.~SI8).
	This is contrary to what reported in Refs.~\onlinecite{Groiss2017,Kuchuk2022}, where the droplet entirely dissolved the GeSn film on its path. This difference can be however explained with the fact that their GeSn film was considerably thinner ($\sim$50nm) compared to the one studied in this work (sample F is 590nm-thick).
	
	Orange symbols in Fig.~\ref{fig_EDX} indicate the positions probed by STEM EDX in Fig.~\ref{fig_EDX}c, with the relative measured atomic compositions reported in the table inset.
	Single elemental maps and additional measurement details are reported in Fig.~SI8.
	As a reference, we consider first the GeSn composition in the pristine region of the monocrystalline GeSn:In film ($\triangle$). Here, STEM EDX detects a Sn concentration of ($5.3\pm0.6$)\%at, close within error to the real GeSn composition measured by XRD. 
	No In is detected in $\triangle$, indicating that its concentration in the film is below the EDX detection limit.
	On the other hand, in correspondence with the segregation droplet (+) STEM EDX reveals major concentrations of Sn and In, respectively of ($91.3\pm1.0$)\%at and ($6.9\pm0.9$)\%at.
	%The two elements are uniformly distributed in the droplet, excluding the presence of phase separation.
	Here, Ge is only detected with a concentration of ($1.8\pm0.2$)\%at, near its solubility limit in pure Sn at the growth temperature of this sample \cite{Olesinski1984}.
	The droplet trail ($\Box$), as expected from previous works \cite{Groiss2017,Kuchuk2022}, is composed of almost pure Ge, with minimal fractions of In and Sn trapped during Ge precipitation from the droplet.
	These observations suggest that  droplet formation is the result of Sn-In co-segregation, as later elaborated in Sec.~\ref{sec:discussion}.

	Having verified that segregation droplets are composed of Sn-In, we can now shift our attention back to Fig.~\ref{fig_Co-SegregationTable}.
	Here, the arrows show that substrate temperature (red), In flux (purple), and Sn flux (yellow) are parameters that can initiate or increment the extent of Sn-In segregation.
	In particular, we can observe the following effects of the fluxes: At a nominal substrate temperature of 205\degree C, samples with a In/Ge flux ratio of 0.05\% (samples A, E) show no sign of segregation. On the other hand, by increasing the In/Ge flux ratio to 0.1\%, we observe segregation for $F_{Sn}/F_{Ge}=10\%$ (sample~D, $x_{Sn}=3.8\%$at), and $F_{Sn}/F_{Ge}=15\%$ (sample~F, $x_{Sn}=5.9\%$at).
	It is remarkable that a small $F_{In}/F_{Ge}$ increase of 0.05\% causes Sn-In segregation, considering that even an increase of 10\% in $F_{Sn}/F_{Ge}$ from sample~A to E does not induce it. Therefore, In seems to have a considerably stronger effect on segregation than Sn, and we will discuss this phenomenon later in Sec.~\ref{subsec:Discussion_CoSegregation}.
	A further increase of $F_{In}/F_{Ge}$ to 0.3\%  induces partial segregation in Ge$_{0.981}$Sn$_{0.019}$ (sample~C), and full segregation in Ge$_{0.943}$Sn$_{0.057}$ (sample~G), clearly visible in the magnified SEM image (G, 40x).
	
	Besides the In flux, we can observe that also the Sn flux has an effect on the extent of segregation: at constant substrate temperature, increasing $F_{Sn}/F_{Ge}$ from 10\%~($x_{Sn}\sim3.8\%$) to 15\%~($x_{Sn}\sim5.7\%$) in presence of $F_{In}/F_{Ge}=0.1\%$ (respectively samples D and F) considerably increases the fraction of surface covered by segregated regions.
	Finally, increasing the substrate temperature (from sample I grown at 185\degree C, to H at 195\degree C, to F at 205\degree C) also induces segregation.
	While the increase of Sn flux and substrate temperature are expected from previous studies to favor segregation in GeSn alloys~\cite{Mukherjee2021,Groiss2017,Zaumseil2018,Kuchuk2022}, the influence of In remains to be elucidated.
	
	\subsection{\label{sec:lowTgrowth}Defective growth at low temperature}
	
	\begin{figure*}[!t]
		\includegraphics[width=\textwidth]{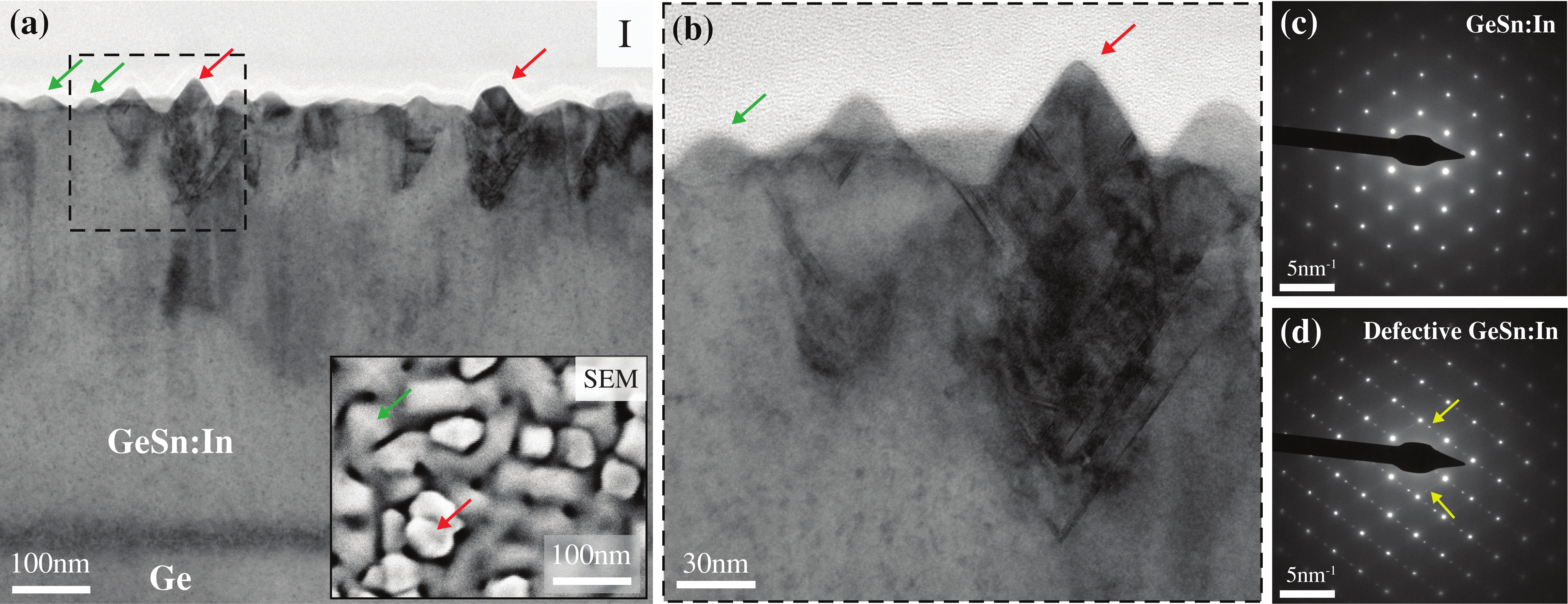}
		\caption{\label{fig_185Cgrowth}(a) Cross-sectional TEM bright-field image of sample~I, grown with a nominal substrate temperature of 185\degree C. The inset shows a top-view SEM image of the sample surface. Red and green arrows mark respectively defective, and non-defective regions.
			(b) Zoomed TEM bright-field image of the region enclosed within the black, dashed square in (a), showing clearly the presence of defects at these regions.
			(c) TEM diffraction pattern of non-defective GeSn:In showing only diffraction spots corresponding to \hkl<110> zone axis.
			(d) TEM diffraction pattern of defective region in (b), showing additional diffraction spots and streaks (yellow arrows) due to the presence of stacking faults.
			Additional details on these measurements in Fig.~SI9.
		}
	\end{figure*} 	
	
	As expected, decreasing the substrate temperature from sample~F can reduce (sample~H) and prevent (sample~I) Sn-In co-segregation.
	This, however, comes at the cost of film crystal quality.
	In Fig.~\ref{fig_Co-SegregationTable}, the 100x zoomed SEM top-view images of non-segregated regions show regularly faceted surface features for substrate temperatures of 195\degree C (H, 100x) and 205\degree C (F, 100x) corresponding to monocrystalline, pseudomorphic GeSn:In.
	On the other hand, in sample~I, grown at a lower $T$ of 185\degree C, the surface morphology is irregular and epitaxy seems broken at the brighter grains.
	Indeed, cross-sectional TEM BF of this sample in Fig.~\ref{fig_185Cgrowth}a shows that pseudomorphic epitaxial growth locally breaks down after the film reaches a thickness of about 400nm, forming defective regions that extend to the film surface.
	A magnified TEM image of one of these defective regions is reported in Fig.~\ref{fig_185Cgrowth}b.
	Here, defect nucleation during growth yields surface asperities (red arrows) that are taller and of  higher aspect ratio compared to the rest of the surface (green arrows), conferring them the bright contrast observed at SEM.
		
	By means of TEM diffraction, we identified the type of defects forming at 185\degree C in Sample~I.
	Fig.~\ref{fig_185Cgrowth}c shows a reference TEM diffraction pattern of the non-defective region of the GeSn:In film.
	Here, only diffraction spots corresponding to monocrystalline GeSn:In aligned along the \hkl<1 1 0> zone axis are visible.
	The TEM diffraction pattern in Fig.~\ref{fig_185Cgrowth}d of the defective region in Fig.~\ref{fig_185Cgrowth}b shows two additional distinctive features indicated by yellow arrows: (1) striking of diffraction spots, and (2) additional diffraction spots positioned between the spots belonging to the \hkl<1 1 0> zone axis.
	While the former originates from the presence of stacking faults, the latter is a more peculiar feature, appearing in our case due to periodically arranged stacking faults.
	A more detailed analysis of these features is out of the scope of this manuscript, and is thus briefly reported in Fig.~SI9.
	In the latter figure, we also show presence of polycrystallinity in sample~I in correspondence with some defects, indicating local breakdown of epitaxy.
	This behavior has been previously observed in MBE growth of GeSn at $T<155$\degree C and was attributed to kinetic roughening effects \cite{Bratland2005,Desjardins1999}.
	Kinetic roughening describes a system where adatom mobility is strongly reduced by the low growth temperatures, leading to significant surface roughening that induces defect nucleation. This results in a switch from monocrystalline to polycrystalline growth, and eventually to amorphous deposition \cite{Xue1993}.
	
	Lastly, we stress that the growth of sample~I has been repeated to confirm the behavior reported in Fig.~\ref{fig_185Cgrowth}, and that the observed defects are not present in pure GeSn grown at 185\degree C (see Fig.~SI3).
	These results therefore suggest that at low growth temperature the presence of In has an additional detrimental effect during epitaxy of GeSn alloys, related to reduced Sn and Ge adatom diffusion \cite{Desjardins1999} and/or accumulation of surfactant-induced defects on surface \cite{Eaglesham1995}.

	\subsection{\label{sec:surfactant}Surfactant effect of In}
\begin{figure}
		\includegraphics[width=0.48\textwidth]{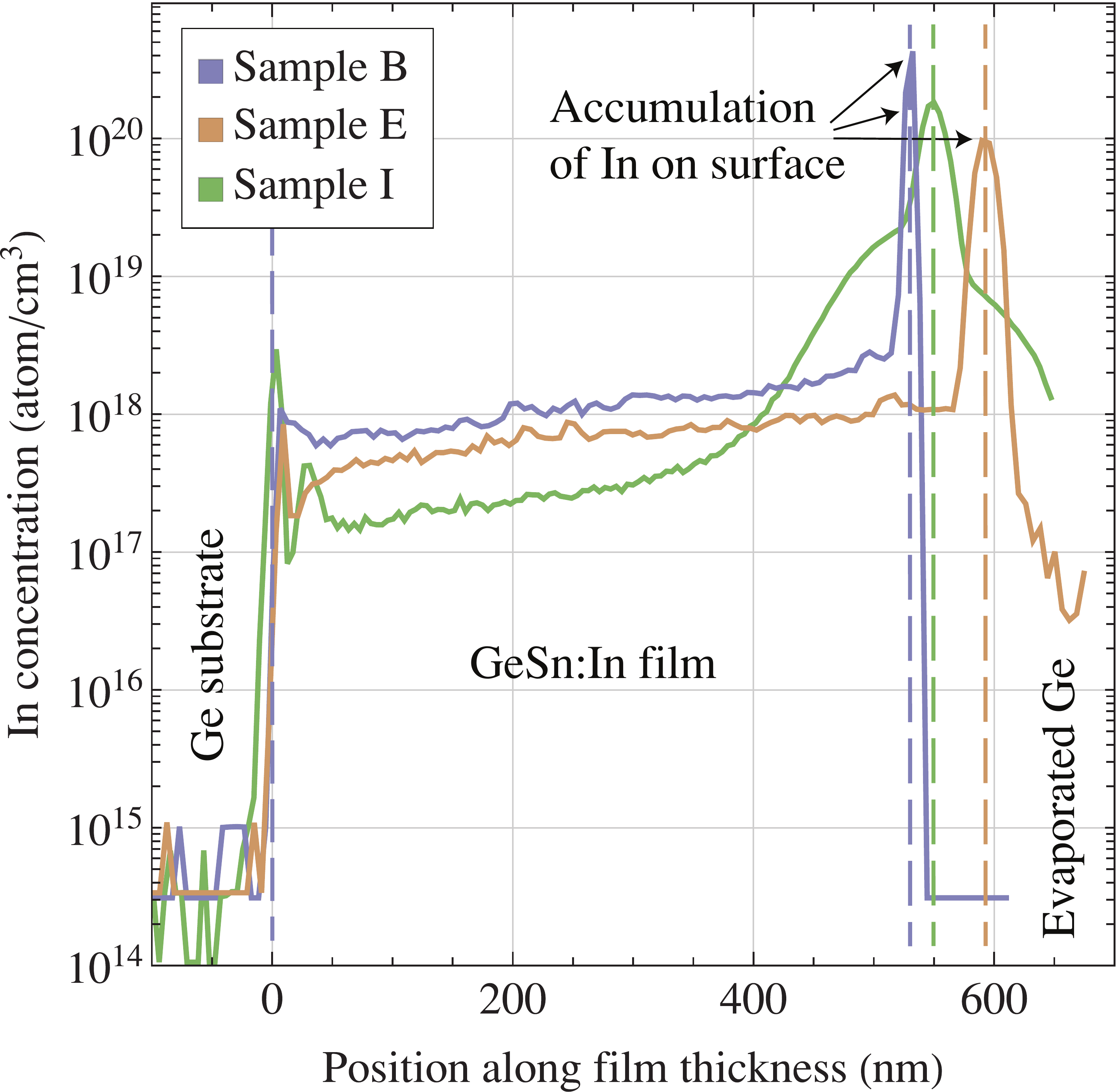}
		\caption{\label{fig_SIMS}
			SIMS depth profiling of In concentration in samples B (purple), E (orange) and I (green). Samples B and E were grown at 205\degree C with different In and Sn fluxes, while sample I was grown at 185\degree C. The three samples were covered by a 100-nm-thick thin film of Ge deposited by evaporation prior to SIMS characterization. SIMS shows that In incorporation increases along the film thickness at all growth temperatures, suggesting that the dopant is acting as a surfactant, accumulating on surface during growth. This is confirmed by the two-order-of-magnitude increase in In concentration on the films surface (indicated by black arrows), which shows clear accumulation of the dopant element on surface. Indium concentration is higher in sample B with respect to sample E as a result of the higher In flux used during growth, while In incorporation seems to be reduced by the lower growth temperature in sample~I. In the latter, the sharp increase of In concentration after 400nm corresponds to the onset of defect nucleation, shown in Fig.~\ref{fig_185Cgrowth}, indicating a higher In incorporation rate in defects.}
	\end{figure}

	To understand whether In acts as a surfactant during growth, aiding surface segregation,
	we performed SIMS depth-profiling of the In concentration in three GeSn:In films.
	The results are shown in Fig.~\ref{fig_SIMS}. We selected samples without surface segregation, namely B (purple), E (orange), and I (green), in order to avoid In concentration artifacts due to Sn-In segregation droplets.
	Samples~B and E were grown at 205\degree C using different Sn and In fluxes, while sample~I was grown at 185\degree C.
	All SIMS profiles show that the In concentration increases along the film thickness, denoting an increment in In incorporation rate as the film grows.
	This indicates that In is acting as a surfactant, accumulating on surface during growth and driving a proportional increase in In incorporation across the entire range of temperatures considered in this study.
	The In surfactant behavior is confirmed by the peaks of two-order-of-magnitude higher In concentration detected at the films surface (indicated by arrows in Fig.~\ref{fig_SIMS}), clearly pointing at an accumulation of the dopant on surface.
	Furthermore, in sample~I we note a strong increase in In concentration already around 400nm, corresponding to the onset of defect nucleation observed in Fig.~\ref{fig_185Cgrowth}.
	This suggests defects can more easily accommodate In dopant incorporation.
	
	\subsection{\label{sec:maxIncorporation}Measured In concentrations in GeSn:In}
	
	By comparing the In atomic concentrations in the different samples in Fig.~\ref{fig_SIMS}, we observe that the In incorporation in the non-defective region of sample~I (up to 400nm) is considerably lower despite its In flux being equal to sample~B and double of sample~E.
	This suggests the In incorporation rate decreases when lowering the growth temperature, possibly due to a reduction of In solubility in Ge \cite{Simoen2007}.
	Furthermore, considering the two samples grown at 205\degree C,  more In has been incorporated in sample B as a result of the higher In flux used during deposition.
	For the same reason, sample~B exhibits a higher In concentration peak on surface, though we should consider that the In surface  peak of sample~E is broadened by its higher film surface roughness (see details in Fig.~SI2).
	Integration of the In surface signal yields an atomic density of $4.0*10^{14}$cm$^{-2}$ and $2.0*10^{14}$cm$^{-2}$, respectively for sample~B and E, corresponding exactly to the ratio between their In fluxes during growth.
	
	From the SIMS depth profiling in Fig.~\ref{fig_SIMS} we can also extract the maximal In incorporation obtained in the non-defective GeSn:In films.
	As a consequence of the In surfactant behavior, the In incorporation rate increases during growth, resulting in a difference in In concentration across the film thickness of almost an order of magnitude. For both samples~B and E the maximal In  concentration is thus found right below the GeSn:In film surface, corresponding respectively to $2.8*10^{18}$cm$^{-3}$ and $1.4*10^{18}$cm$^{-3}$.
	The ratio of maximal In concentrations in the two samples matches the ratio of In fluxes used during their growths, suggesting a direct proportionality between the In flux and the In incorporation rate despite the surfactant behavior of the dopant element.
	It is interesting to notice that both concentrations are lower than the solid solubility of In in pure Ge, reported to be $\sim4*10^{18}$cm$^{-3}$~\cite{Simoen2007}.
	In addition, Hall measurements of sample~B, reported in Fig.~SI6, determined an electrically active carrier concentration of $2.9*10^{17}$cm$^{-3}$ at 300K, yielding a low activation of 24.2\% with respect to the average film In concentration of $1.2*10^{18}$cm$^{-3}$.
	Post-growth thermal annealing may be beneficial in increasing dopant activation, though the process would be limited by the metastability of the system, and is thus preferably avoided in \emph{in-situ} doping \cite{Vohra2019,Wang2016}.
	
	\section{\label{sec:discussion}Discussion}
	
	\subsection{\label{subsec:Discussion_CoSegregation}Origins of Sn-In co-segregating behavior}
	The results of this study indicate a tendency for Sn and In to co-segregate, forming liquid droplets on surface.
	During growth, these droplets move around on surface, dissolving a portion of the GeSn:In film at their front and depositing behind a trail of almost pure Ge, as elucidated in Refs.~\onlinecite{Groiss2017,Kuchuk2022} for the pure GeSn alloy.
	In this section, we provide an explanation for the observed enhancement of segregation induced by In doping.
	
\begin{figure*}
		\includegraphics[width=\textwidth]{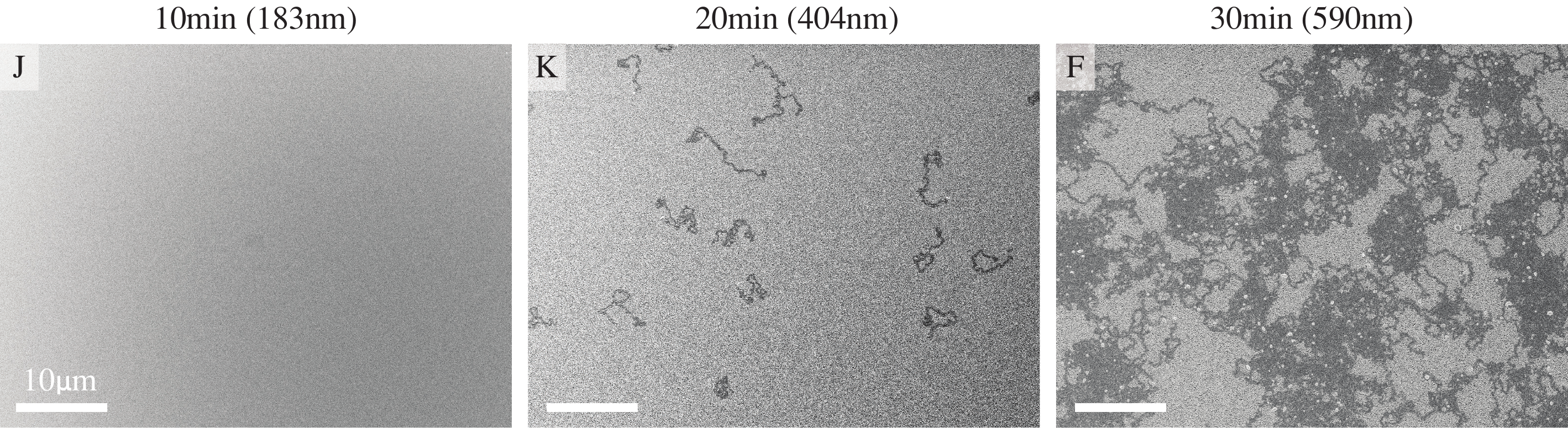}
		\caption{\label{fig_TimeSeries} SEM top-view imaging of the surface of three GeSn:In films (samples J, K, F) grown with the same deposition parameters for different times. Surface segregation is absent after 10min (183nm), while after 20min (404nm), few Sn-In droplets with single trails can be observed. The accumulation of In surfactant on surface during growth induces more extended segregation after 30min (590nm) of growth.}
	\end{figure*}

	Contrary to what observed by Shimura~\emph{et al.} in Ref.~\cite{Shimura2012} for Ga, we found that In does not lose its surfactant properties in presence of Sn at the temperatures considered in this study.
	Therefore, In tends to stay on surface and accumulate during growth as a result of an unbalance between incoming atomic flux and incorporation rate in the film.
	Intuitively, as the film grows, and the In surface concentration increases, we can expect this increase in In to be determinant in initiating segregation.
	To demonstrate it, we grew GeSn:In films with the same deposition parameters as sample F, interrupting the growth to observe the evolution of surface segregation at different film thicknesses. The SEM top-view images of samples J, K and F, respectively grown for 10min, 20min and 30min are shown in Fig.~\ref{fig_TimeSeries}.
	While no segregation occurs with 10min~(183nm) of growth, surface segregation is initiated after 20min~(404nm), and after 30min~(590nm) approximately half of the surface is covered by Sn-In droplets and their trails.
	If there was no increase in In surface concentration during growth, conditions would be stationary, and thus the film thickness would not affect the extent of surface segregation.
	Hence, these results clearly point at In surface accumulation being a driver for Sn-In segregation.
	
	Nevertheless, the In surfactant behavior does not offer a full picture of the segregation process.
	At this point, it is still not clear why Sn and In co-segregate.
	This matter is easily resolved by looking at the InSn phase diagram~\cite{Massalski1990}, which predicts that at the growth temperatures used in this work Sn and In are liquid, and completely miscible for alloy compositions up to 80\%at of Sn. Sn and In can thus aggregate in a single liquid phase, though some excess Sn may precipitate during the process.
	Secondly, the peak value of In concentration reached on surface of sample B measured by SIMS in Fig.~\ref{fig_SIMS} is~$5.8*10^{20}$at/cm$^3$ (see Fig.~SI5 for details on the fitting of the SIMS depth profile), which is approximately 1.2\%at of the cubic Ge atomic density of $4.41*10^{22}$at/cm$^3$.
	Sample~B, with 2\%at Sn, is not segregated. However, with the same In flux, by increasing the Sn fraction to 3.8\%at (sample D) segregation occurs (see note~\footnote{\label{fn_SI-SIMS}
		Here, we assume that the In incorporation rate is independent on the Sn flux, and thus the In surface concentrations of samples~B and D are equal prior to segregation. Fig.~\ref{fig_SIMS_SI} confirms this is the case, by showing that the In concentrations below the surface of samples~A and E are practically equal, despite their different Sn flux.} in this regard).
	Summing the Sn and In fractions in sample D we obtain 5\%at, which is well below the fraction of Sn necessary to cause segregation in intrinsic GeSn at this growth temperature (see e.g. sample E, with 5.7\%at Sn, is not segregated, despite being doped with a small concentration of In).
	Simply put, while 5\%at Sn does not cause segregation at a nominal substrate temperature of 205\degree C, 3.8\%at Sn plus 1.2\%at In do.
	This indicates that the increase in segregation in presence of In is not due to a mere increase in total concentration of segregating atoms. It is rather the nature of the In+Sn system that is increasing the occurrence of surface segregation when replacing 1.2\%at Sn with 1.2\%at In (see note~\footnote{The present calculation of surface concentration of 1.2\%at In is clearly a simplification, but it is justified by looking at the composition of the segregation droplet in Fig.~\ref{fig_EDX} (sample~F), where the ratio of atomic densities $n_{In}/n_{Sn}$ is equal to 7.6\%.
	Additional SEM EDX measurements on different SnIn droplets in Fig.~\ref{SI_DropletsSEMEDX} confirmed the ratio $n_{In}/n_{Sn}$ lies between 5\% and 10\% in sample~F.
	Given the almost full miscibility of the SnIn liquid phase in this range of alloy composition \cite{Massalski1990}, we can fairly assume that this ratio represents the ratio of surface adatom concentrations of the two elements on surface.
	In the case of sample~F, with 5.9\%at Sn, this ratio thus yields a surface composition of $\sim(0.45\pm0.15)$\%at In, which is of the same order of magnitude of the 1.2\%at In obtained from SIMS measurements on sample~B, grown under an equal In flux.
	This validates our initial argument on the enhancement of segregation due to the presence of In, especially considering that 0.45\%at$<$1.2\%at implies an even stronger influence of In on the segregation behavior of GeSn.}).
	
	To explain the influence of In, we turn to classical theory of nucleation and growth of thin films~\cite{Pound1954,Venables1984}.
	Droplet surface segregation can be described as the 3D nucleation of a liquid phase on the film surface.
	%Surface nucleation is a process governed by thermodynamics, with two conflicting contributions: a positive contribution of the enthalpy of formation of the droplet phase ($g_V$, molar Gibbs free energy of formation), and a negative contribution from the increase in surface area related to the formation of the nucleus ($\sigma_v$, surface tension of the droplet). The formation of a new surface
	The nucleation rate ($J$) is then expressed with the scaling law (using the notation from Ref.~\onlinecite{Venables1984})
	\begin{equation}
		\label{eq_NucleationRateJ}
		J\propto D n_1 n_i
	\end{equation}
	where $n_1$ is the surface concentration of single adatoms, $D$ is their diffusion coefficient, and $n_i$ is the concentration of critically-sized clusters.
	We expand the latter using
	\begin{equation}
		\label{eq_CriticalClusterConc}
		n_i\propto\exp[-\Delta G_i/(k_BT)]
	\end{equation}
	where $k_B$ is the Boltzmann constant, $T$ is the substrate temperature, and $\Delta G_i$ is the Gibbs free energy of a cluster at its critical size, which can be approximated by
	\begin{equation}
		\label{eq_GibbsCriticalCluster}
		\Delta G_i \propto X^3/\Delta\mu^2
	\end{equation}
	with $X$ being the liquid surface tension, and $\Delta\mu$ the supersaturation of the liquid phase.
	Substituting eqs.~\ref{eq_CriticalClusterConc} and \ref{eq_GibbsCriticalCluster} in eq.~\ref{eq_NucleationRateJ} we obtain
	\begin{equation}
		\label{eq_NucleationRateJSubst}
		J\propto D n_1 \exp[-X^3/(\Delta\mu^2k_BT)]
	\end{equation}
	With eq.~\ref{eq_NucleationRateJSubst} in mind, we can compare a system with 5\%at Sn on surface to a system with 3.8\%at Sn plus 1.2\%at In, at equal temperature. We know the former system does not segregate, while the latter does.
	The total adatom concentration $n_1$ would be the same for both systems, and is therefore not playing a role in the observed difference in segregation.
	On the other hand, the diffusion coefficient $D$ is different for the two atomic species. In literature, we find diffusion energy barriers for In and Sn adatoms on the Si(100) surface, respectively equal to 0.27eV~\cite{Albao2009}, and 1.2eV~\cite{Dolbak2010}. Due to the similarity of group-IV materials, we can expect analogous behavior on the surface of a Ge(100) substrate and GeSn(100) film.
	The considerably smaller diffusion barrier for In signifies In adatoms can diffuse faster than Sn adatoms.
	Faster diffusion increases the adatom diffusion length, and thus its likelihood of encountering and sticking to an atomic cluster.
	This ultimately results in greater probability of droplet nucleation. 
	
	A further element that may enhance segregation in presence of In is the liquid surface tension $X$. Dadashev \emph{et al.} report in Ref.~\onlinecite{Dadashev2019} that the surface tension of the SnIn liquid is smaller by 1-2\% (depending on the In atomic fraction) compared to pure Sn liquid, lowering the Gibbs free energy of critical clusters ($\Delta G_i$, eq.~\ref{eq_GibbsCriticalCluster}).
	In Ref.~\onlinecite{Dadashev2019}, the study included a range of temperatures between 250\degree C and 450\degree C, reasonably close to the growth temperatures used in this work to expect an analogous behavior  of $X_{SnIn}$.
	Lastly, also the supersaturation term, inversely proportional to the equilibrium vapor pressure of the species ($\Delta\mu=k_BT\ln(p/p_e)$, with $p$ being the partial pressure of an element, and $p_e$ its equilibrium vapor pressure), predicts an increase in segregation in presence of In: Firstly,  $p_{e,In}<p_{e,Sn}$~\cite{Gray1972}, and secondly, considering the only available data in the literature, the activity coefficient (see note~\footnote{\label{fn_activityCoefficient}For more information on the meaning of \emph{activity coefficient}, the reader is directed to the brief, but complete review of Ref.~\onlinecite{Tomiska1998}}) of In in the SnIn melt at 400\degree C is well below 1~\cite{Kumari2008,Hultgren1973}, suggesting a further decrease in $p_{e,In}$. The activity coefficient of Sn is very close to 1 for large Sn fractions in the SnIn liquid~\cite{Kumari2008,Hultgren1973}, and should thus not play a role here.
	Overall, the supersaturation term $\Delta\mu$ is thus likely increased in magnitude in presence of In at our growth temperatures, driving an increase in the nucleation rate $J$.
	To summarize, at a fixed GeSn:In growth temperature, the terms $D$, $X$, and $\Delta\mu$ from eq.~\ref{eq_NucleationRateJSubst} hint at a higher liquid droplet nucleation rate in presence of In, explaining the increase in segregation with respect to pure GeSn epitaxy.
	
	Concerning the  dependence of Sn-In co-segregation with temperature and Sn flux observed in Fig.~\ref{fig_Co-SegregationTable}, SIMS measurements in Fig.~\ref{fig_SIMS} show that within the range of growth parameters used in this work In always acts as a surfactant, inducing Sn-In segregation. In addition, from eq.~\ref{eq_NucleationRateJSubst} we can deduce that
	\begin{itemize}
		\item A lower Sn flux trivially decreases $n_1$ and $p_{Sn}$, and thus the nucleation rate $J$.
		\item The Arrhenius-type dependence of nucleation rate $J$ on the temperature $T$ determines a decrease in $J$ with lower $T$.
		\item The same holds for $D$, which is also dependent on $T$ via an Arrhenius-type law.
		\item The surface tension $X$ is weakly dependent on $T$ \cite{Dadashev2019}, and increases with lower $T$, lowering $J$.
		\item $p_{e}$ decreases with lower $T$ and is thus the only term positively contributing to $J$. Experimental data in Fig.~\ref{fig_Co-SegregationTable} however clearly shows that co-segregation is reduced by lowering the substrate $T$, and thus the positive contribution from $p_{e}$ must be lower than the negative contribution of all other terms of eq.~\ref{eq_NucleationRateJSubst}.
	\end{itemize}
	
	Lastly, we highlight that the present analysis focused on a narrow range of growth temperatures, and that it is unclear whether In would favor segregation also below this range. Still, the results from growth at 185\degree C in Fig.~\ref{fig_185Cgrowth} showed that the range of usable growth temperatures for GeSn:In is limited by the presence of In itself. Even if Sn-In do not co-segregate at lower temperatures, defects will breakdown epitaxy, preventing low-temperature monocrystalline GeSn:In growth.

	\subsection{\label{subsec:Discussion_InSolubility}In solubility in GeSn}
	From Fig.~\ref{fig_SIMS}, the maximal In incorporation we obtained was $2.8*10^{18}$cm$^{-3}$ in Ge$_{0.98}$Sn$_{0.02}$ (sample~B), which is lower than the solubility of In in pure Ge  ($\sim4*10^{18}$cm$^{-3}$~\cite{Simoen2007}).
	All other growth parameters equal, an increase in Sn content to 3.8\%at (sample D) and 5.7\%at (sample F) leads to progressively increased Sn-In segregation.
	Hence, we can conclude that the solubility of In in non-defective GeSn thin films is dependent on the alloy composition, and decreases with increasing Sn content.
	The solubility of In is thus expected to be lower in GeSn compared to pure Ge.
	
	This dependence of In incorporation on alloy composition could arise from the compressive strain present in the pseudomorphic GeSn:In films considered in this study. As reported in Tab.~\ref{tab_SampleDescriptions}, compressive strain increases with Sn fraction in the alloy, disfavoring the insertion of large Indium atoms in the GeSn matrix.
	Further experiments are required to verify if In solubility is dependent on the strain state of the thin film or if it is purely dependent on the alloy composition.
	
	\subsection{\label{subsec:Discussion_InMaxIncorporation}Maximizing In incorporation in GeSn}
	Considering the maximal dopant incorporation measured in this work, we note that it is substantially smaller than the maximal GeSn \emph{in-situ p-type} active doping concentrations $>10^{20}$cm$^{-3}$ reported in the literature~\cite{Wang2016,Vohra2019}, especially considering the low electrical dopant activation of 24.2\% measured in sample~B.
	In principle, the In concentration of Ge$_{0.98}$Sn$_{0.02}$ in this sample could be increased by increasing the In flux, but in Fig.~\ref{fig_Co-SegregationTable} sample~C shows that by triplicating $F_{In}$, the Ge$_{0.98}$Sn$_{0.02}$:In film segregates.
	This yields $8.4*10^{18}$cm$^{-3}$ as upper limit of In incorporation in non-defective GeSn at 205\degree C, still considerably lower than the concentrations of \emph{p-type} dopants found in literature.
	
	Typically, dopant incorporation could be increased by kinetically hindering Sn-In co-segregation, i.e. lowering the growth temperature.
	However, the growth of sample~I at 185\degree C showed that also this possibility is limited, as the In incorporation rate decreases at lower temperature (Fig.~\ref{fig_SIMS}). In addition, at the same temperature, epitaxial growth of GeSn:In starts breaking down after 400nm (Fig.~\ref{fig_185Cgrowth}), while this does not occur for pure GeSn.
	Maintaining a GeSn:In thickness lower than 400nm would certainly prevent the nucleation of defects at 185\degree C, but a further decrease in temperature would still be limited by the kinetic roughening effect, since the critical thickness for epitaxial breakdown is known to be reduced by lowering the growth temperature \cite{Nerding2003}.
	Furthermore, when employing thinner GeSn:In films, one needs nonetheless to consider that the dopant incorporation will be limited by the surfactant behavior of In, which determines a difference in concentration of almost one order of magnitude across film thicknesses of less than 600nm, as visible in Fig.~\ref{fig_SIMS} for samples~B and E.
	%In addition, a further decrease in growth temperature to maximize the In flux, while preventing segregation would decrease the critical thickness for epitaxial breakdown \cite{Nerding2003}, limiting the range of growth temperatures for GeSn:In even with films thinner than sample I.
	%We attribute this behavior to the surfactant effect of In, which may have two effects: (I) In decreases the surface diffusion of Ge adatoms , resulting in kinetic roughening, similar to what observed when excessively lowering the Ge growth temperature \cite{Bratland2005}, (II) In passivates island edges \cite{Kandel2000} favoring islanding in low epitaxial mismatch growth.
	%Link to figure in pure Ge in SI
	
	Overall, this study does not aim at maximizing the In doping concentrations, but rather at elucidating the limitations associated with \emph{in-situ} In doping of GeSn. Still, the phenomena we outlined show that the practical maximal In doping concentration in non-defective GeSn:In is not too far from the values measured in our work.
	
	% pROBALY SEGREGATION OCCURS WHEN IN SOLUBILITY IS REACHED, BECAUSE AT THESE GROWTH TEMEPRATURE IT IS TOO MOBILE, AND IT PROBABLY FOLLOWS EQUILIBRIUM BEHAVIOR.
	
	\section{\label{sec:conclusions}Conclusions}
	The results shown in this work outline some limitations for using In as \emph{p-type} dopant element in GeSn.
	We demonstrated that In induces co-segregation with Sn, causing the formation of mobile SnIn metallic droplets on the GeSn:In film surface that could be detrimental for (opto)electronic device performance.
	We illustrated the enhancement in segregation as the result of multiple factors.
	First, In acts as a surfactant for GeSn, accumulating on surface during growth.
	Secondly, In adatoms diffuse faster than Sn adatoms, increasing the probability to encounter segregating clusters and bond to form stable liquid nuclei.
	Thirdly, the limited data present in the literature suggest that the Gibbs free energy of formation of a critical SnIn liquid nucleus is lower compared to that of pure Sn liquid.	
	
	We observed that as a result of the surfactant effect of In, its incorporation rate increases during growth, complicating accurate control of the final dopant concentration in the film.
	The maximum dopant incorporation we measured in non-defective films was $2.8*10^{18}$cm$^{-3}$ in Ge$_{0.98}$Sn$_{0.02}$ with a low electrical activation of 24.2\% at 300K, far from the maximal active \emph{p-type} dopant values $>10^{20}$cm$^{-3}$ reported in the literature for GeSn. 
	Furthermore, we showed that the solubility of In in GeSn decreases with larger Sn fractions in the alloy, limiting the applications of GeSn:In \emph{in-situ} doping to devices that do not require significant doping concentrations.
	Lastly, we demonstrated that lowering the growth temperature to avoid Sn-In co-segregation and push the In incorporation is not a viable strategy for this material system, as it leads to a decrease in In incorporation rate and epitaxy breakdown due to kinetic roughening effects and/or dopant-related defect accumulation.
	
	This work provides new insights on the behavior of the Indium dopant element in the GeSn system, and discourages its utilization in GeSn-based optoelectronic devices.
	
	\begin{acknowledgments}
		The authors wish to thank Nihal Singh for his help in substrate preparation, and Jean-Baptiste Leran for his maintenance of the MBE. This work was supported by Innosuisse, SNSF NCCR QSIT, Max Planck Institut für Festkörperforschung, and Max Planck Graduate Center for Quantum Materials.
		
		\textbf{Author contributions}: A.G. and A.F.M. conceived the experiments. A.G. did all the experiments and analysis with the exception of Hall measurements in Fig.~SI6, performed by L.E.W., and atomic force microscopy measurements in Fig.~SI2, performed by T.H. The manuscript was written by A.G., with inputs from A.F.M and L.E.W.
	\end{acknowledgments}

	\bibliography{GeSnIn_manuscript}
		
	\end{document}

	% Use the \preprint command to place your local institutional report number 
	% on the title page in preprint mode.
	% Multiple \preprint commands are allowed.
	%\preprint{}
	
	\title[]{Supporting Information for: Surfactant behavior and limited incorporation of Indium during in-situ doping of GeSn grown by MBE}
	
	% Force line breaks with \\
	\author{A. Giunto}
	\author{L. E. Webb}
	\author{T. Hagger}
	\author{A. Fontcuberta i Morral}
	\email{anna.fontcuberta-morral@epfl.ch}
	\affiliation{ 
		Laboratory of Semiconductor Materials, Institute of Materials, École Polytechnique Fédérale de Lausanne, Lausanne, Switzerland
		%\\This line break forced with \textbackslash\textbackslash
	}%
	
	% repeat the \author .. \affiliation  etc. as needed
	% \email, \thanks, \homepage, \altaffiliation all apply to the current author.
	% Explanatory text should go in the []'s, 
	% actual e-mail address or url should go in the {}'s for \email and \homepage.
	% Please use the appropriate macro for the type of information
	
	% \affiliation command applies to all authors since the last \affiliation command. 
	% The \affiliation command should follow the other information.
	
	%\author{}
	%\email[]{Your e-mail address}
	%\homepage[]{Your web page}
	%\thanks{}
	%\altaffiliation{}
	%\affiliation{}
	
	% Collaboration name, if desired (requires use of superscriptaddress option in \documentclass). 
	% \noaffiliation is required (may also be used with the \author command).
	%\collaboration{}
	%\noaffiliation
	
	\date{\today}

	\maketitle %\maketitle must follow title, authors, abstract and \pacs
	
	\renewcommand{\thefigure}{SI\arabic{figure}}
	\setcounter{figure}{0}
	
	\renewcommand{\thetable}{SI\arabic{table}}
	\setcounter{table}{0}

		\begin{table}[h!]
		\centering
		\caption{\label{tab_SampleDescriptionsSI}MBE deposition parameters of Ge$_{1-x}$Sn$_x$:In films studied in this work. Fluxes are reported in \emph{nTorr}, as per measurement from the MBE beam flux monitor. Alloy composition and (compressive) strain are measured by XRD RSM, shown in Fig.~\ref{figSI_RSM}.
		Substrate nominal temperatures were calibrated with an infrared thermal camera, yielding errors of $\pm20$\degree C associated with the substrate transparency and variable emissivity over the IR camera wavelengths range of 7.5\textmu m to 13\textmu m. We also report nominal $T$ measured with a thermocouple (TC) standing behind the substrate holder, \textbf{not} in contact with it.
		}
		\begin{tabular*}{0.8\textwidth}{c @{\extracolsep{\fill}}|c |c |c |c |c |c |c |c |c |c}
			\makecell{Sample\\ label} &  \makecell{Sub. T\\ (\degree C)} & 
			\makecell{TC T\\ (\degree C)} & \makecell{Ge flux \\ (nTorr)} &  \makecell{Sn flux \\ (nTorr)} &  \makecell{In flux \\ (nTorr)} &  \makecell{Gr. time \\ (min)} &  \makecell{Thickness \\ (nm)} & \makecell{$x_{Sn}$ \\ (\%)}  & \makecell{Degree of \\ relaxation} & \makecell{In-plane \\ strain (\%)}\\
			\hline
			A & 205 & 260 & 1000 & 50 & 0.5 & 30 & 547 & 1.8  & 0&-0.26 \\
			\hline
			B & 205 & 260 & 1000 & 50 & 1 & 30 & 542 & 2.0  & 0&-0.30 \\
			\hline
			C & 205 & 260 & 1000 & 50 & 3 & 30 & 552 & 1.7 & 0 & -0.80%\textsuperscript{\ref{foot_PartialSegregation}}
			\\
			\hline
			D & 205 & 260 & 1000 & 100 & 1 & 30 & 565 & 3.8 &0& -0.56%\textsuperscript{\ref{foot_PartialSegregation}}  & 
			\\
			\hline
			E & 205 & 260 & 1000 & 150 & 0.5 & 30 & 592 & 5.7 & 0&-0.82\\
			\hline
			F & 205 & 260 & 1000 & 150 & 1 & 30 & 590 & 5.9 &0& -0.85%\footnote{\label{foot_PartialSegregation}Sn is partially segregated out of the film, reducing the incorporated Sn content with respect to non-segregated films with the same Ge-Sn flux ratio.}
			\\
			\hline
			G\footnote{\label{foot_FullSegregation}Sn is mostly segregated out of the film. Only two weakly intense peaks remain, which identify partially relaxed GeSn.} & 205 & 260 & 1000 & 150 & 3 & 30 & 571 & \makecell{2.3\textsuperscript{\ref{foot_FullSegregation}}\\ 5.7\textsuperscript{\ref{foot_FullSegregation}}} & \makecell{0.13\textsuperscript{\ref{foot_FullSegregation}}\\ 0.26\textsuperscript{\ref{foot_FullSegregation}}}
			& \makecell{-0.11\textsuperscript{\ref{foot_FullSegregation}}\\ -0.63\textsuperscript{\ref{foot_FullSegregation}}} \\
			\hline
			H & 195 & 250 & 1000 & 150 & 1 & 30 & 580 & 5.3 &0& -0.78%\textsuperscript{\ref{foot_PartialSegregation}}  
			\\
			\hline
			I & 185 & 240 & 1000 & 150 & 1 & 30 & 583 & 5.4 & 0&-0.80 \\
			\hline
			J & 205 & 260 & 1000 & 150 & 1 & 10 & 183 & 5.7 & 0&-0.85\\
			\hline
			K & 205 & 260 & 1000 & 150 & 1 & 20 & 404 & 5.4 &0& -0.80%\textsuperscript{\ref{foot_PartialSegregation}}
			%\\
			%\hline
			%L & 170 & 1150 & 92 & - & 30 & 480 & 6.0 & - & -
		\end{tabular*}
	\end{table}

		\begin{figure*}[h!]
			\caption{\label{figSI_RSM}Reciprocal space maps (RSM) of the (224) planes for the samples grown in this work.
			The Ge substrate peak position is at coordinates \{5.000,~7.075\}nm$^{-1}$, while the GeSn peaks shift position depending on the alloy composition, following Vegard's law~[C. Xu et al., J. Appl. Phys. 122 (2017) 125702.].
			RSM characterisation shows that all GeSn:In films are pseudomorphic, compressively strained, since their in-plane lattice parameter, measured by $Q_x$, is equal to that of the Ge substrate. 
			\emph{Sample J} presents an anomalous RSM due to defects in the original Ge substrate.
			It is however clear that the grown GeSn:In film in \emph{sample J} is pseudomorphic, since every point in the RSM of the Ge substrate has an equally distanced RSM point along the same $Q_x$ coordinate, corresponding to the GeSn epitaxial film. This allowed to calculate the Sn fraction reported in Tab.~\ref{tab_SampleDescriptionsSI}, just like for the other samples.}
			\includegraphics[width=0.43\textwidth]{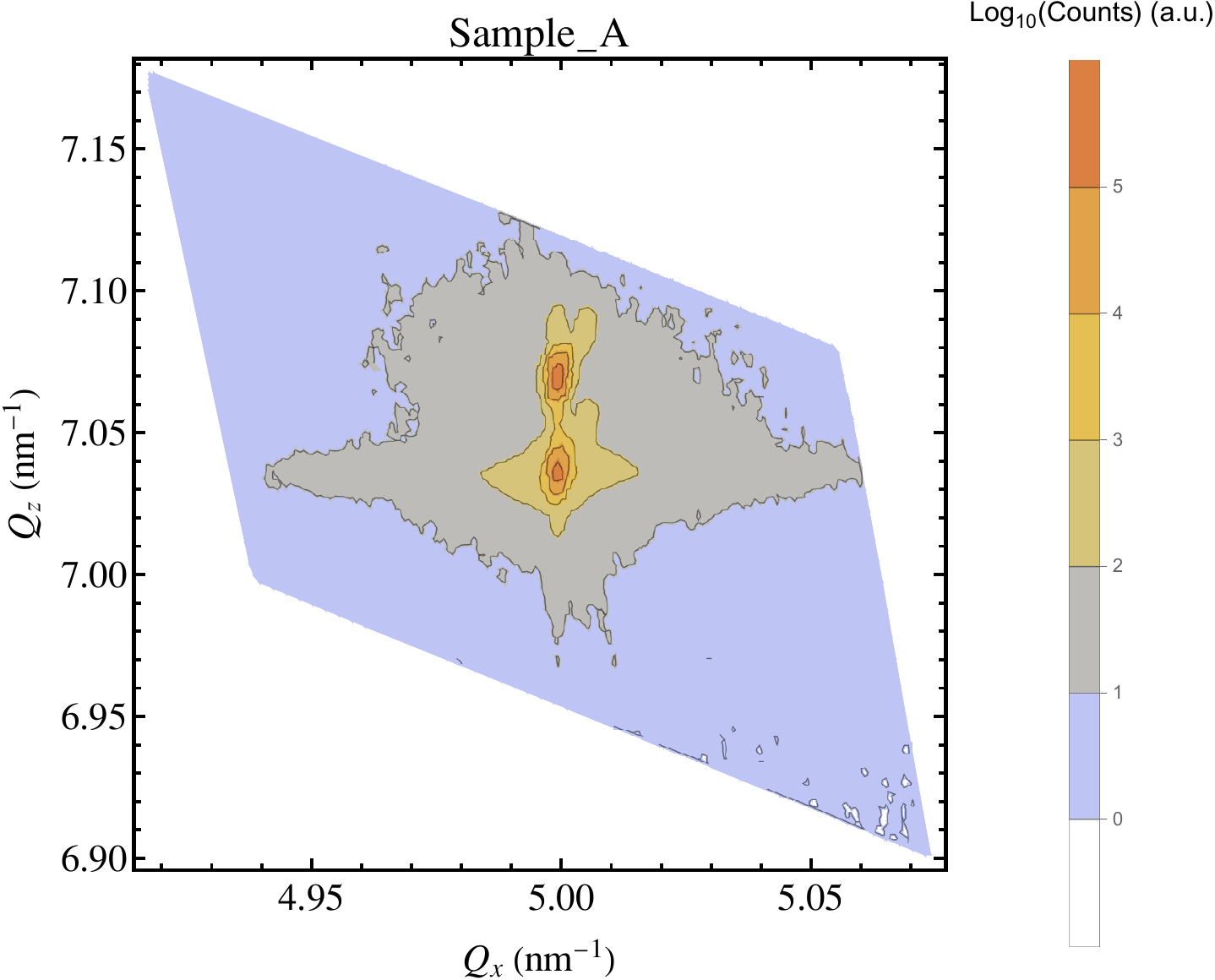}\hfill
			\includegraphics[width=0.43\textwidth]{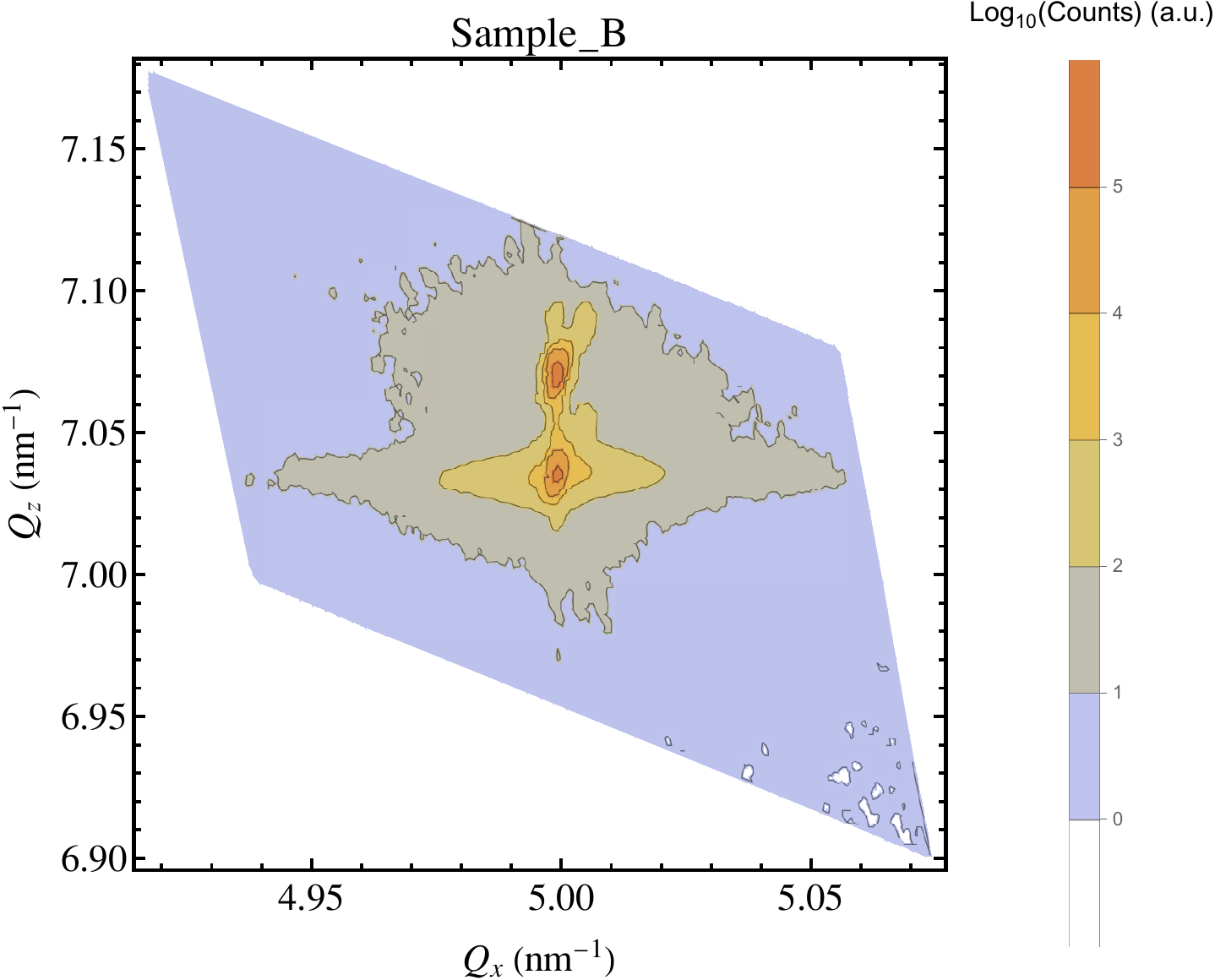}\hfill
		\end{figure*}
	
		\begin{figure*}[h!]
			%\ContinuedFloat
			\includegraphics[width=0.49\textwidth]{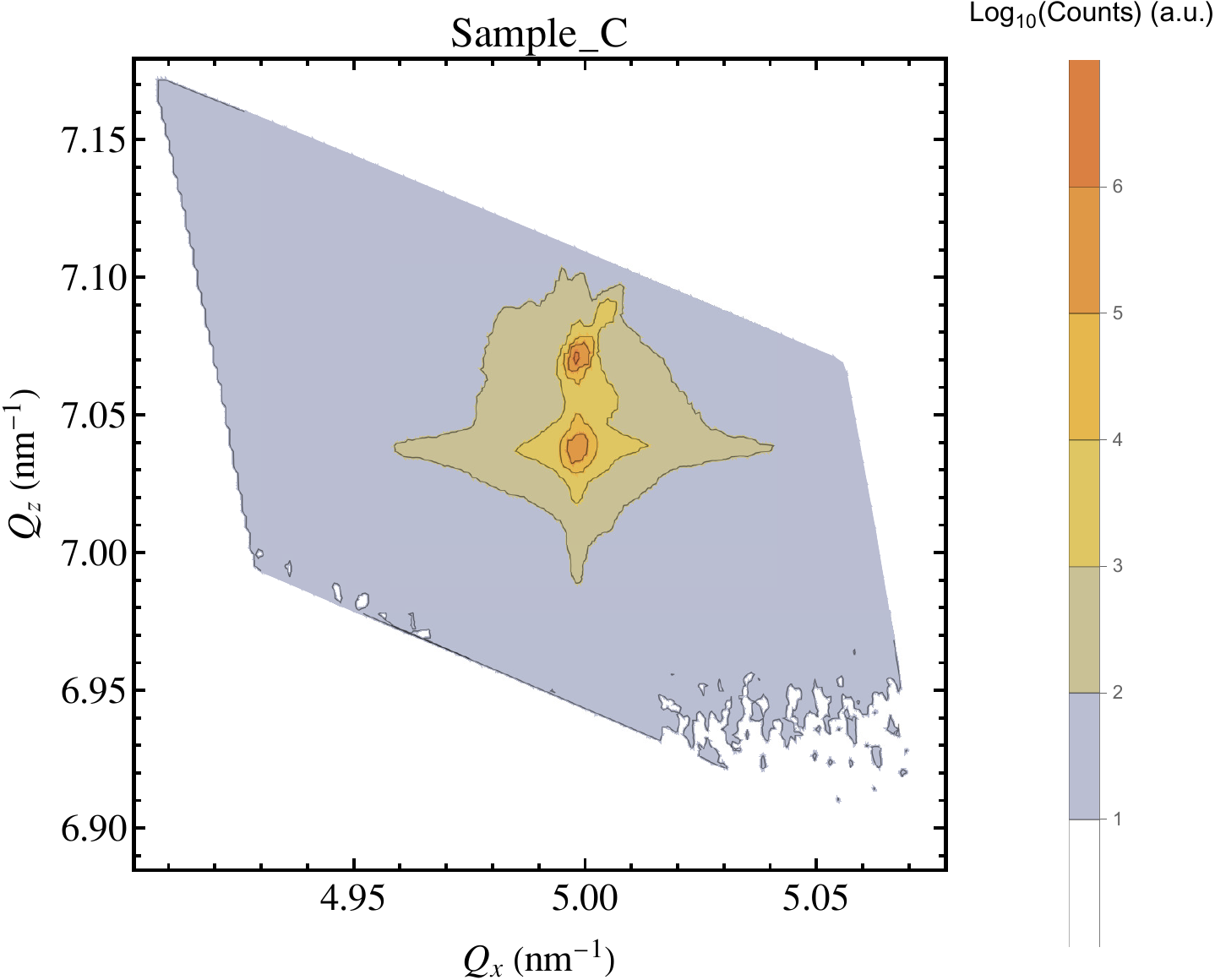}\hfill
			\includegraphics[width=0.49\textwidth]{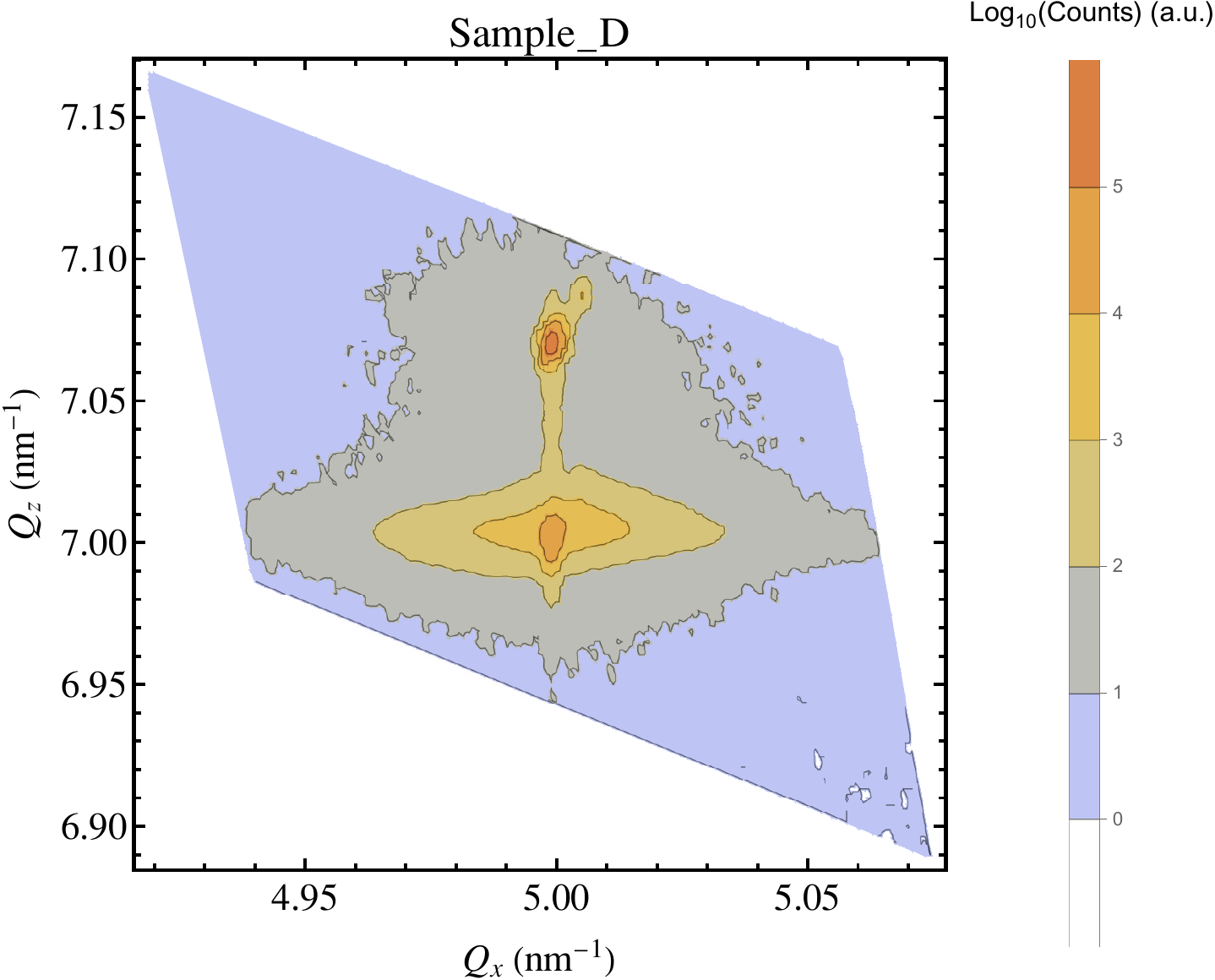}\hfill
			\\[\smallskipamount]
			\includegraphics[width=0.49\textwidth]{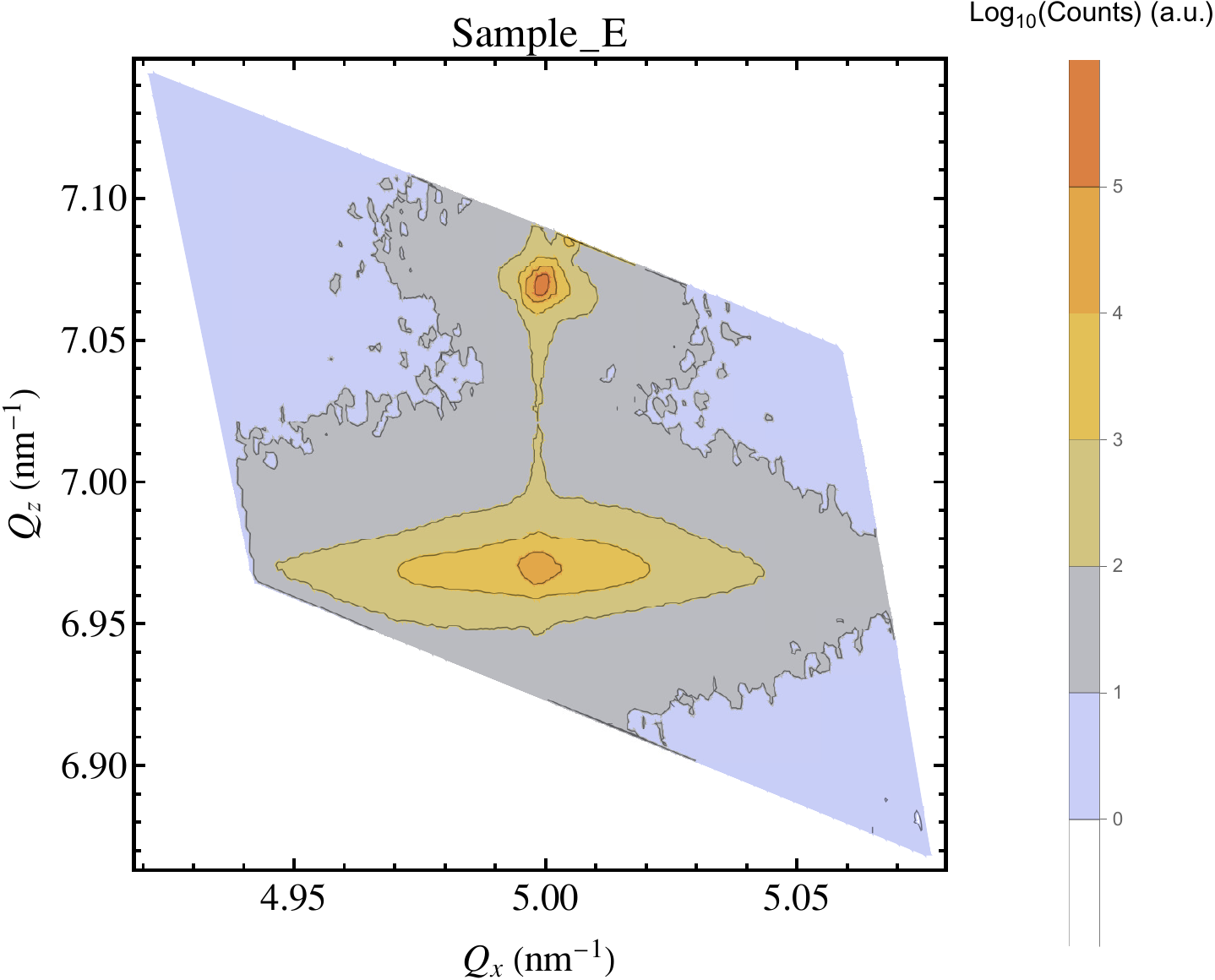}\hfill
			\includegraphics[width=0.49\textwidth]{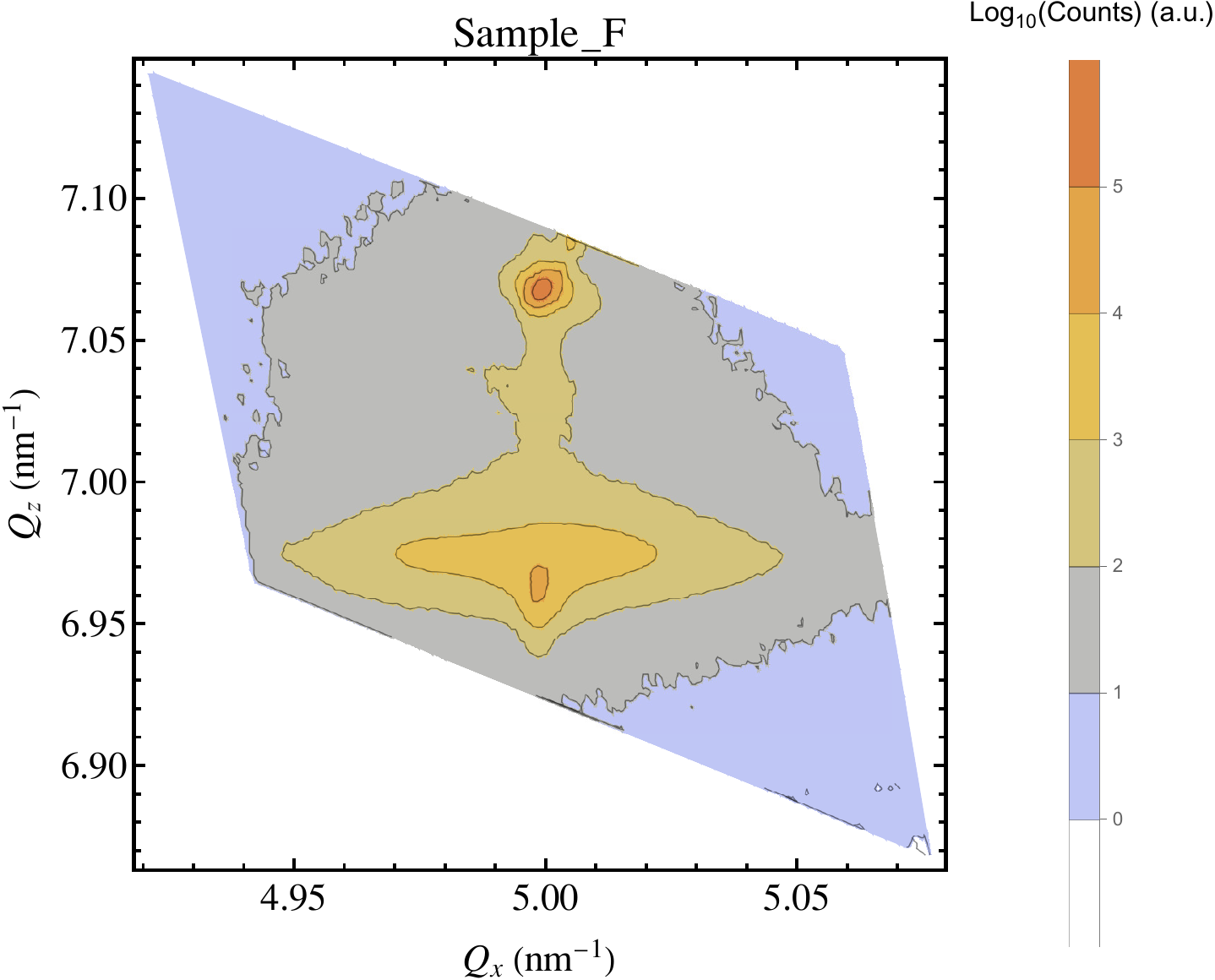}\hfill
			\\[\smallskipamount]
			\includegraphics[width=0.49\textwidth]{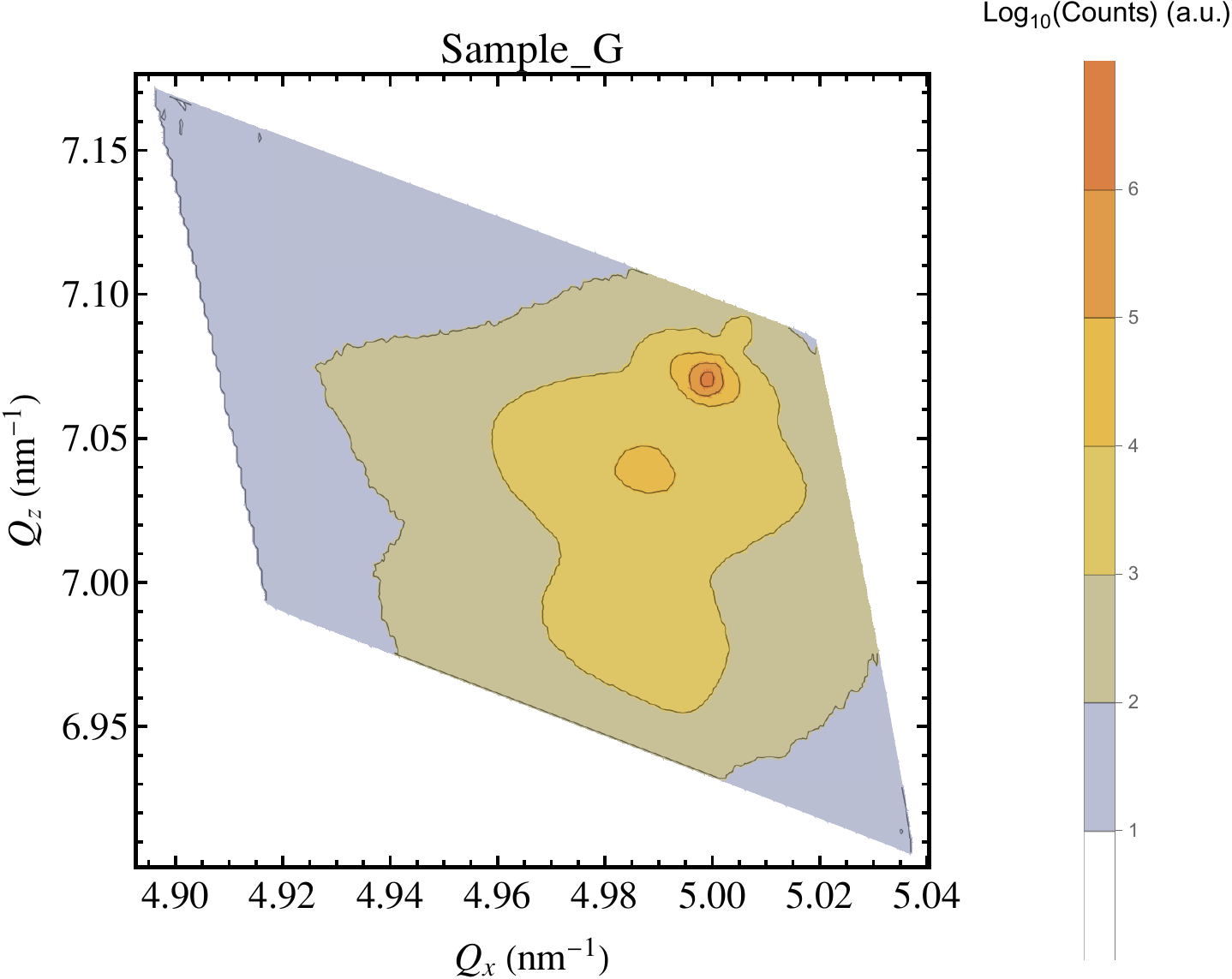}\hfill
			\includegraphics[width=0.49\textwidth]{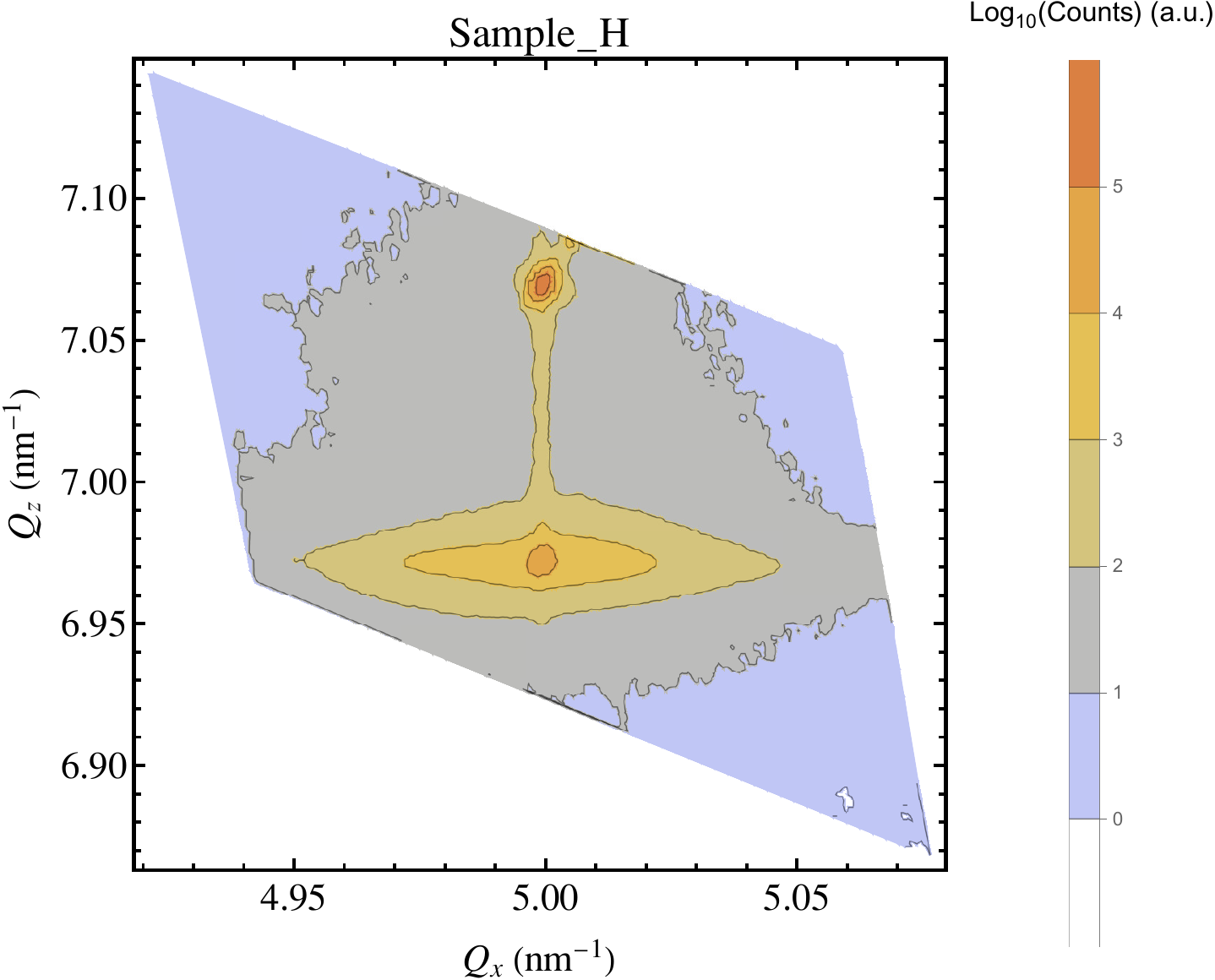}\hfill
		\end{figure*}

		\begin{figure*}[h!]
			\includegraphics[width=0.49\textwidth]{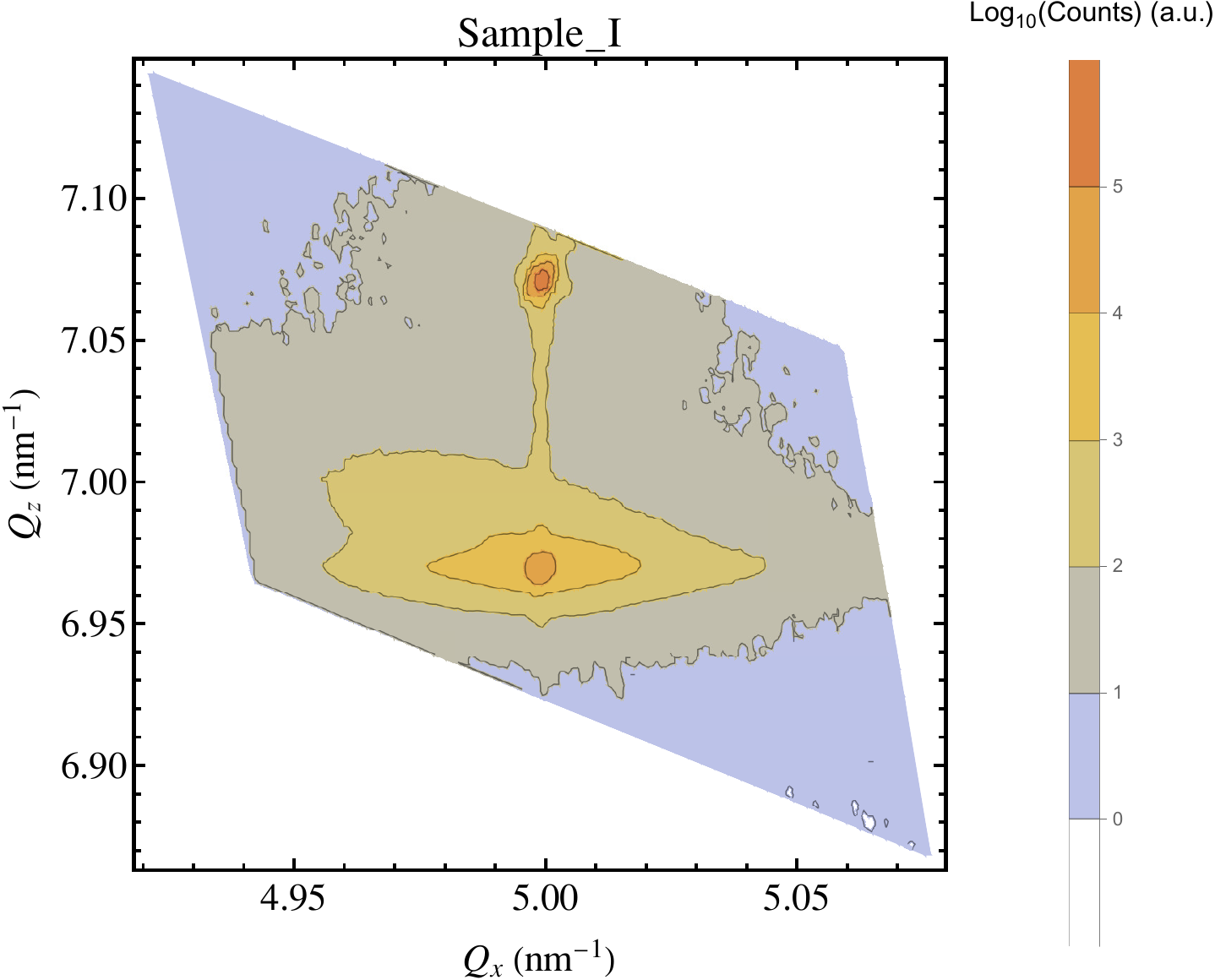}\hfill
			\includegraphics[width=0.49\textwidth]{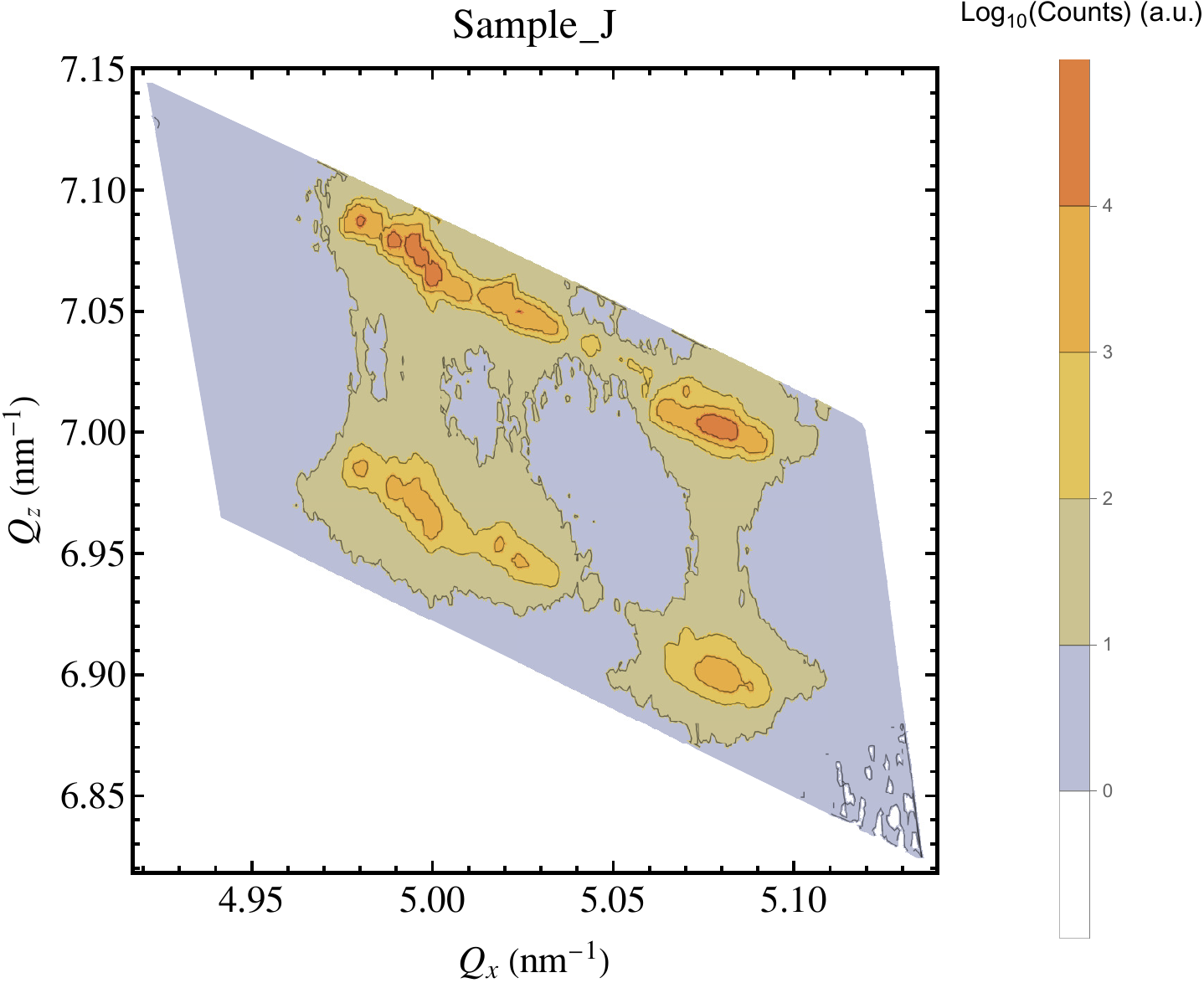}\hfill
			\\[\smallskipamount]
			\includegraphics[width=0.49\textwidth]{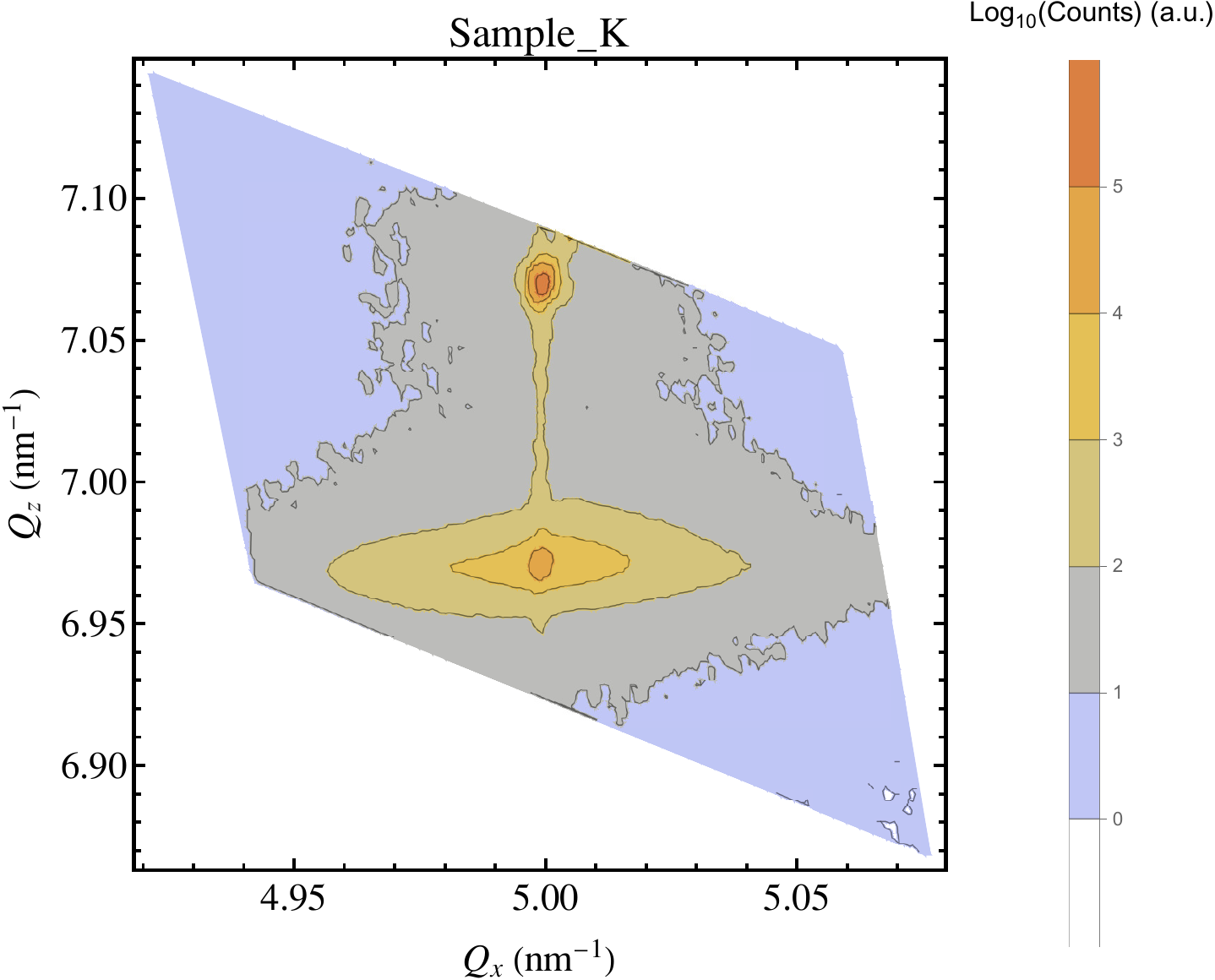}\hfill
		\end{figure*}
	
		\begin{figure}[h!]
			\caption{\label{fig_SI-AFM} Atomic force microscopy (AFM, \emph{Bruker FastScan AFM}) characterisation of the surface of samples~B and E. The measured RSM roughness is respectively 2.4nm and 7.6nm, and the z-range (difference in height between highest and lowest point) is respectively 17nm and 46nm.
			The z-ranges measured here correspond almost perfectly to the width of the In surface SIMS signals of the two samples.
			The large roughness and z-range of sample~E therefore explain the broadening artifact observed in its SIMS measurement of In surface concentration.}
			\includegraphics[width=0.7\textwidth]{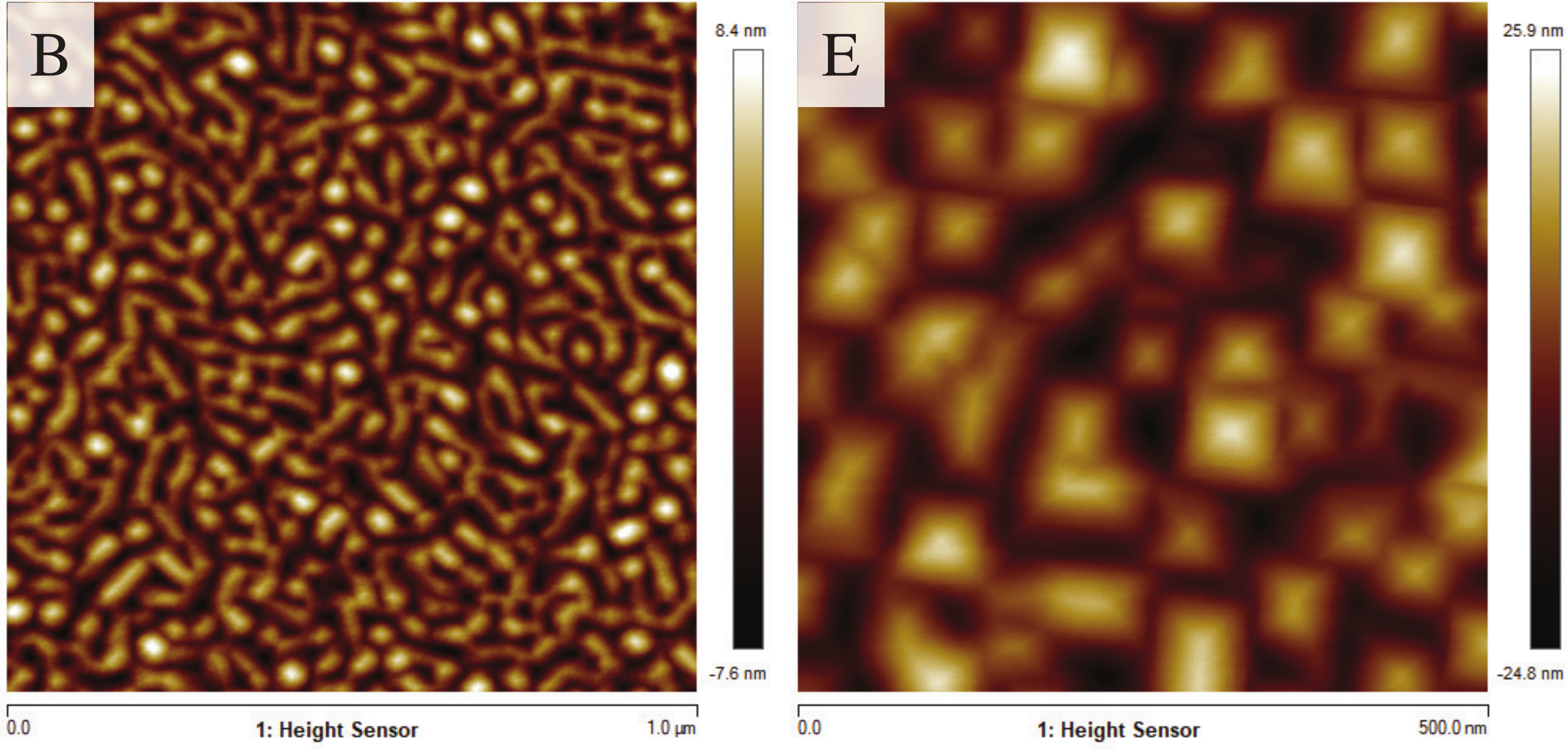}
		\end{figure}
	
		\begin{figure}[h!]
			\caption{\label{fig_SI-pureGeSnLowT} SEM top-view images of a pure Ge$_{0.94}$Sn$_{0.06}$ film grown at 170\degree C, with no In flux. This sample, with a thickness well above 400nm, has surface features similar to samples F and H, visible in the magnified SEM images in Fig.~1, or sample E, visible in the AFM in Fig.~\ref{fig_SI-AFM}. Despite the low growth temperature of 170\degree C, the surface of this pure GeSn film is not defective. This shows that the defects present at 185\degree C on the surface of sample I are due to the presence of In, and not merely due to its low growth temperature.}
			\includegraphics[width=0.49\textwidth]{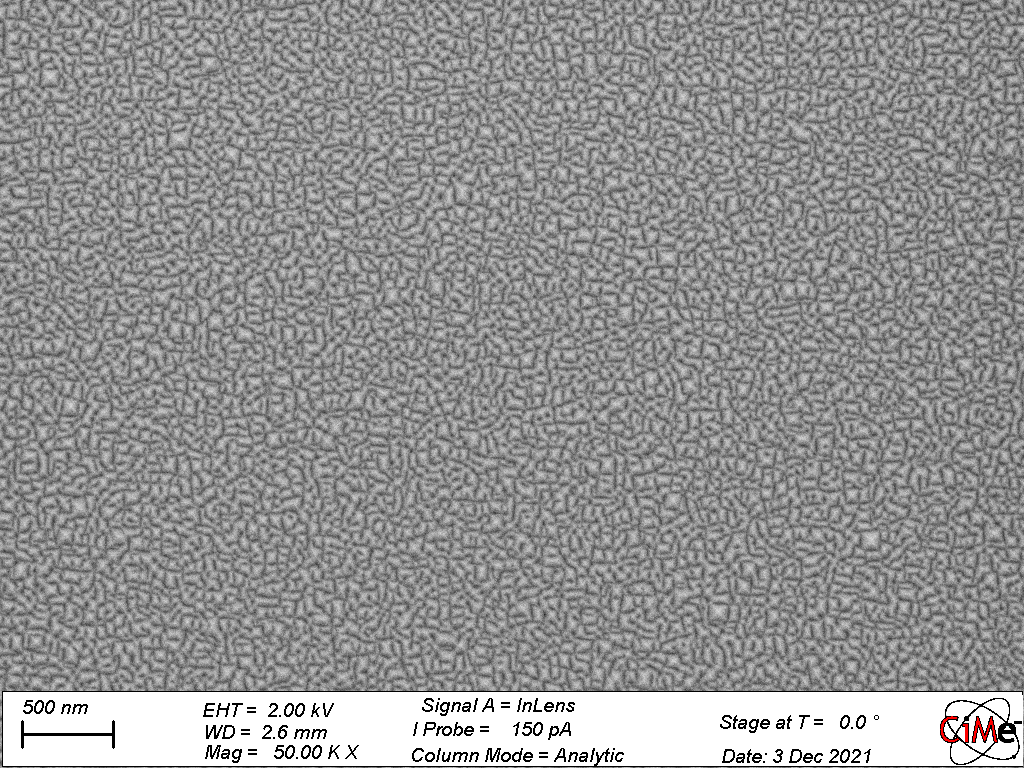}
			\hfill
			\includegraphics[width=0.49\textwidth]{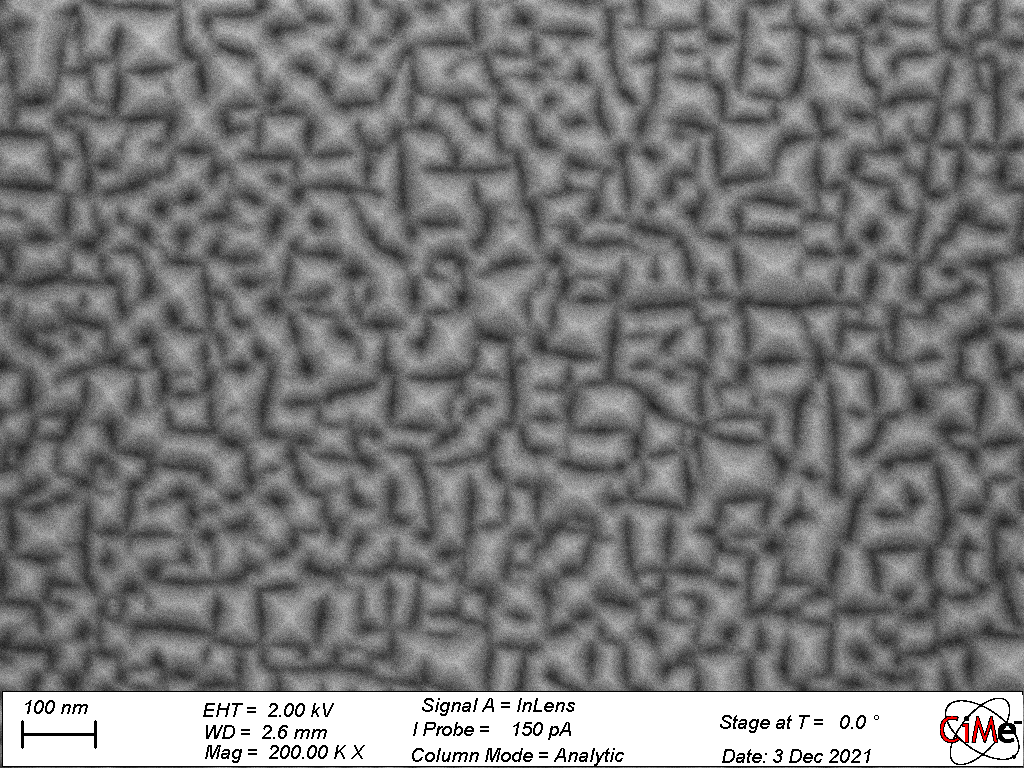}
		\end{figure}
	
			\begin{figure}[h!]
			\caption{\label{fig_SIMS_Sn_SI} SIMS Sn depth profiling of samples B, E, I, and A.
				%Sn composition values on the y-axis are obtained by calibrating the SIMS signal (in a.u.) to match the alloy composition measured by XRD RSM.
				SIMS shows that the composition of the GeSn alloy is uniform in the film, as the Sn signal remains constant across the GeSn:In thickness.
				The zoomed region shows that a slight accumulation of Sn is present on surface, especially at lower Sn contents in samples B and A, due to residual adatoms that were not incorporated at the end of the film growth.
				This intensity of Sn on the film surface corresponds at most to 2.5 times the Sn film signal, and is therefore not comparable to the accumulation of In due to surfactant effect, since in the latter case we have an increase of In concentration of more than 2 orders of magnitude.
			}
			\includegraphics[width=\textwidth]{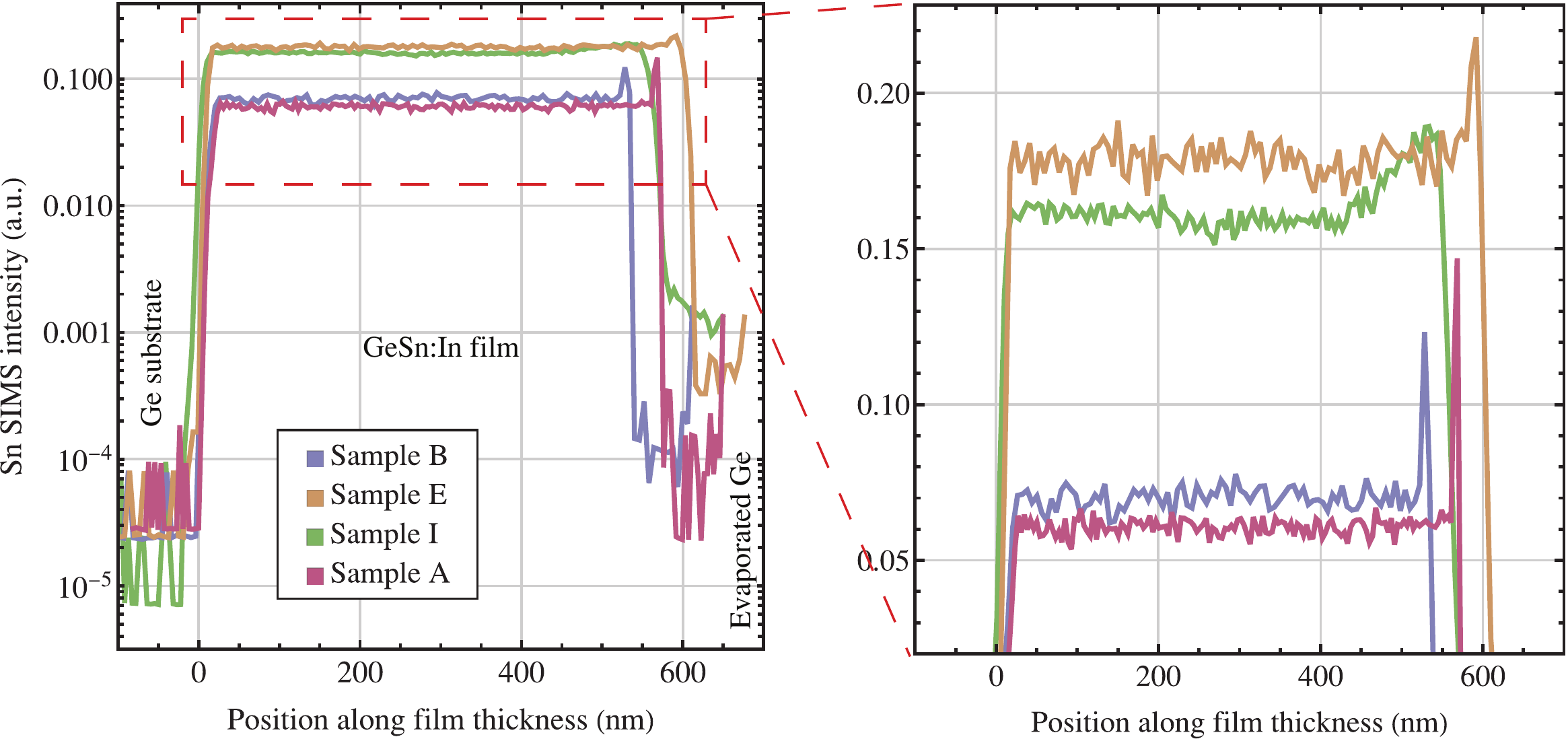}
		\end{figure}	
	
		\begin{figure}[h!]
			\caption{\label{fig_SIMS_SI} (a)~SIMS depth profiling of Fig.~4, with the addition of sample A.
			SIMS shows that the In incorporation in samples~E and A right below the surface is practically equal, suggesting that we can consider In incorporation to be independent on the Sn flux. (b)~SIMS In surface peaks of samples~B and E, with relative Gaussian fitting used to estimate the peak In concentration. This was found to be $5.8*10^{20}$at/cm$^3$ and $4.7*10^{20}$at/cm$^3$ respectively for samples~B and A. The latter has lower In concentration due to the lower In flux during deposition.
			}
			\includegraphics[width=0.6\textwidth]{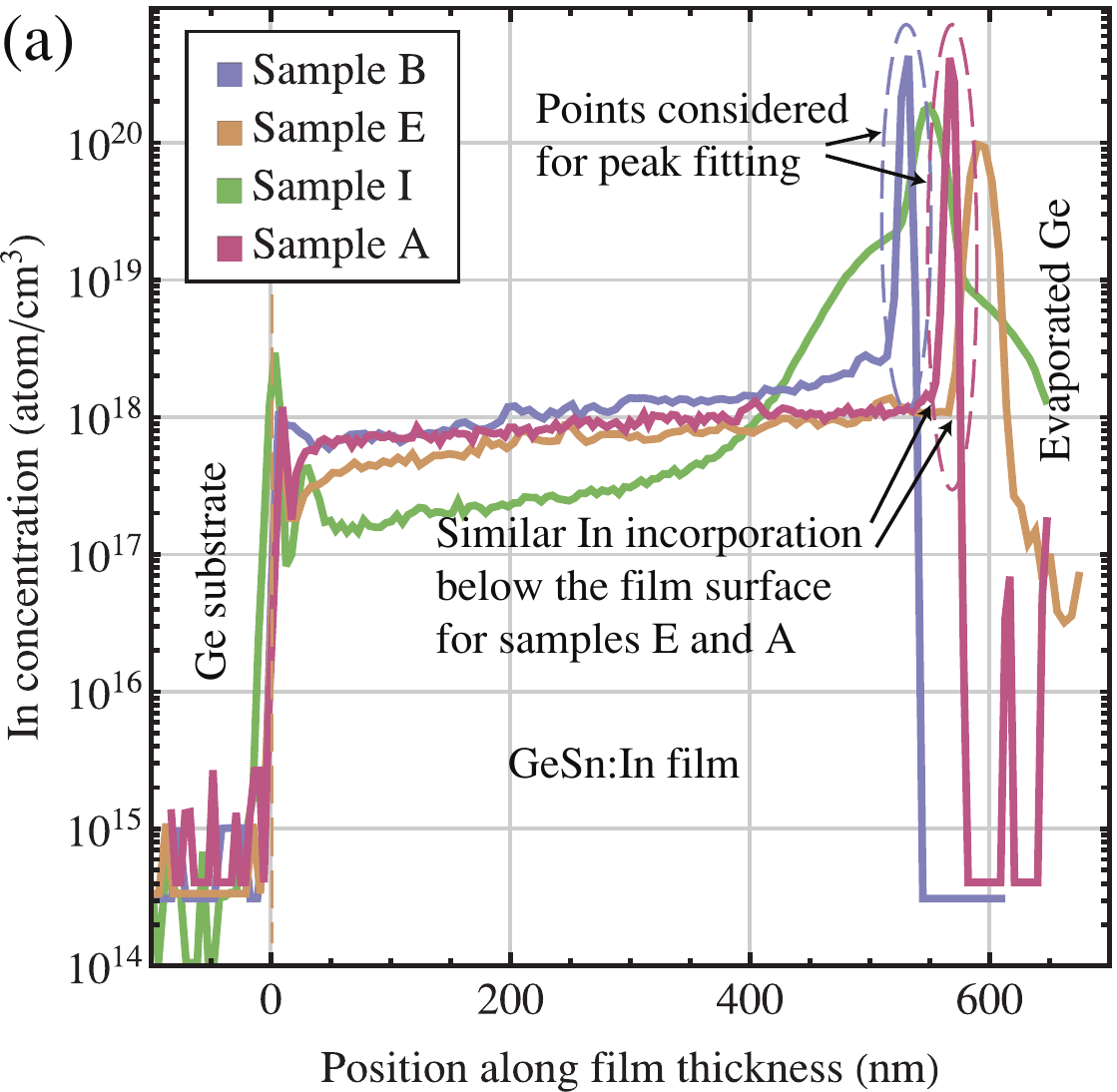}
			\\[\bigskipamount]
			\includegraphics[width=0.6\textwidth]{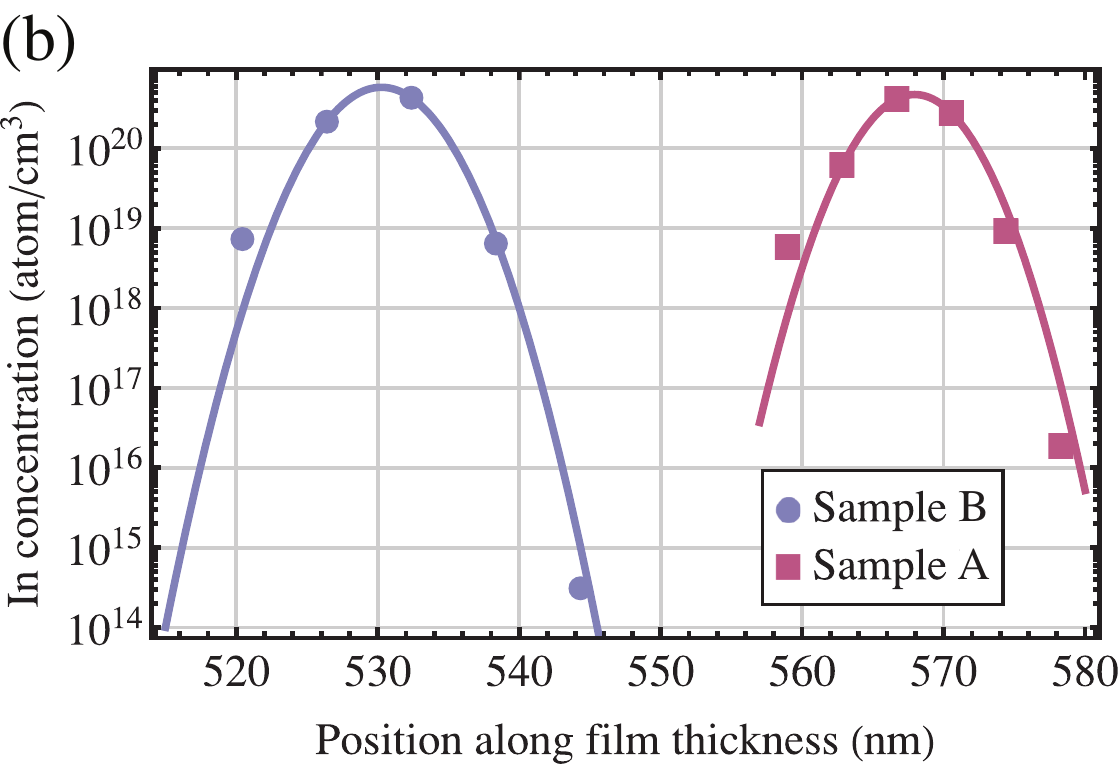}
		\end{figure}

		\FloatBarrier %To place text after figure
		\begin{figure}[h!]

			\caption[]{\label{figSI_HallsampleB} (a) Hall resistance and (b) carrier concentration and mobility measured in sample~B. The sample was diced into a chip of 7mm~x~7mm, and triangle-shaped, 300-nm-thick Al contacts were evaporated in the corners in van der Pauw configuration, with 50-nm-thick Ti adhesion layers.
				The measurements were performed in a \emph{Quantum Design} Physical Property Measurement System between 300K and 150K.}
			\includegraphics[height=0.42\textwidth]{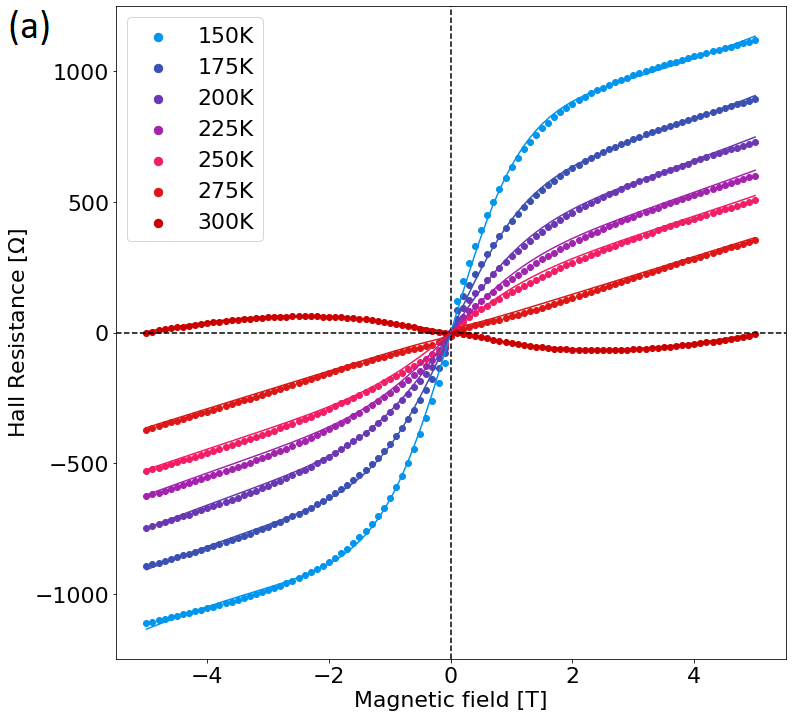}\hfill
			\includegraphics[height=0.42\textwidth]{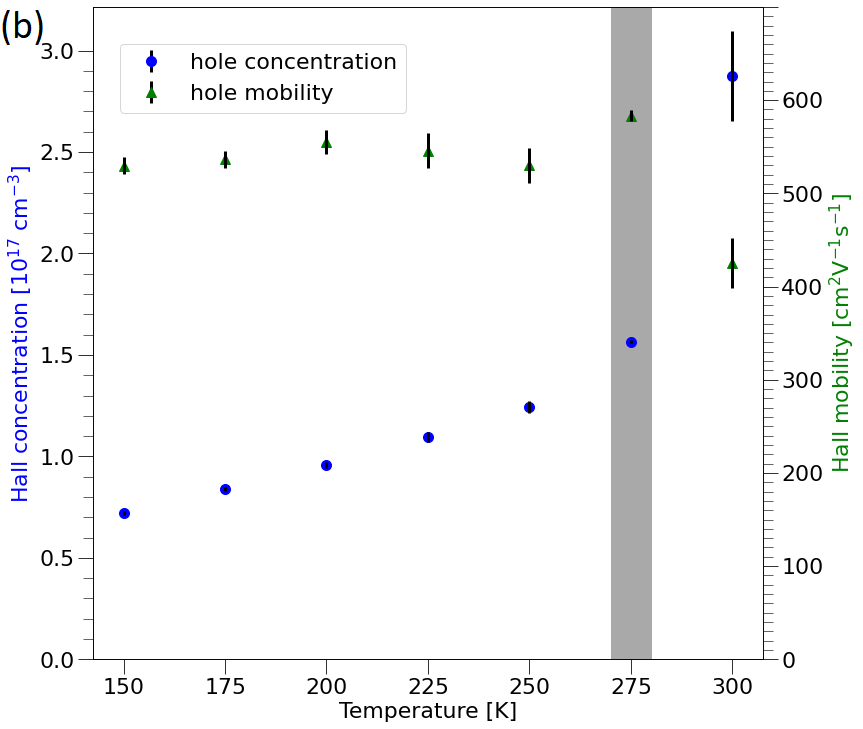}\hfill
		\end{figure}
			
			The shape of Hall resistance in Fig.~\ref{figSI_HallsampleB}a shows that the Ge(001) substrate contributes to the signal [N. Sircar \emph{et al.}, Phys. Rev. B. 83 (2011) 125306.].
			To isolate the film electrical properties, we fitted the Hall resistance data using a two-layer conduction model. This model treats the GeSn:In film and Ge substrate as parallel conductors and predicts the magnetic field $B$ dependence of the Hall resistance $R_{xy}$ to be given by [G. Pettinari \emph{et al.}, Appl. Phys. Lett. 101 (2012) 222103.]:   
			
			\begin{displaymath}
				R_{xy}(B)=\frac{B}{e} \frac{ d_1 n_1 \mu_1^2 + d_2 n_2 \mu_2^2+( d_1 n_1 + d_2 n_2) \mu_1^2 \mu_2^2 B^2 }{(d_1  |n_1| \mu_1^2 + d_2 |n_2| \mu_2^2)^2 + ( d_1 n_1 + d_2 n_2)^2 \mu_1^2 \mu_2^2 B^2  }
			\end{displaymath}
			where $e$ is the elementary charge, $d_1,d_2$ are the film and substrate thicknesses, $n_1, n_2$ are the carrier concentrations and $\mu_1, \mu_2$ are the carrier mobilities.  The subscripts $1$ and $2$ refer respectively to the properties of the film and substrate.
			To model electrons in the film or substrate, negative signs are applied to $n_1$ or $n_2$ respectively, in agreement with the negative sign on the Hall coefficient when electrons are the dominant carrier.
			As in the work from Pettinari \emph{et al.}, to reduce the number of parameters and avoid overfitting we employed the measured zero-field longitudinal resistivity ($\rho_{xx}(T)$; curve not shown here) as boundary condition, using the following formula:
			
			\begin{displaymath}
				\rho_{xx}(T,B=0)= \frac{d_1 + d_2}{e(d_1 |n_1| \mu_1 + d_2 |n_2| \mu_2 )}
			\end{displaymath}

			For temperatures between 150K and 250K, the best fit for the Hall resistance data in (a) was found by assuming p-type conduction in both the film and substrate layers. On the other hand, at 300K, the best fit was provided by considering p-type conduction in the GeSn:In film and n-type conduction in the Ge substrate, with carrier concentration and mobility similar to the intrinsic values at 300K.
			These results indicate that at temperatures below 250K, p-type parallel conduction arises in the Ge substrate due to the presence of acceptor impurities. This is reasonable, as our Ge(001) substrates have a nominal  resistivity  of $>$30Ohmcm at 300K, implying the possible presence of impurities that reduce the substrate resistivity compared to the intrinsic value of 46Ohmcm.
			At 300K, we measure intrinsic conduction in the Ge substrate as n-type because the mobility of electrons is considerably higher than that of holes.
			The intrinsic hole concentration still contributes to the measurements though, resulting in larger error bars at 300K compared to $T\leq250$K.
			
			On a side note, although the model appears to fit well at 275K, at this temperature the carrier concentration and mobility values calculated in the substrate have very large associated error, suggesting the two-layer parallel conduction model does not capture the full picture.
			Indeed, at $T=275K$, ambipolar conduction occurs in the Ge substrate, complicating the fitting. With these conditions, a more complex model is required to extract accurate values of carrier concentration and mobility in the sample.

		\begin{figure}[h!]
			\caption{\label{figSI_SEM_zoom}Zoomed SEM images of sample F, with different magnifications (2kx, 5kx, 10kx, 25kx, 50kx).}
			\includegraphics[width=0.49\textwidth]{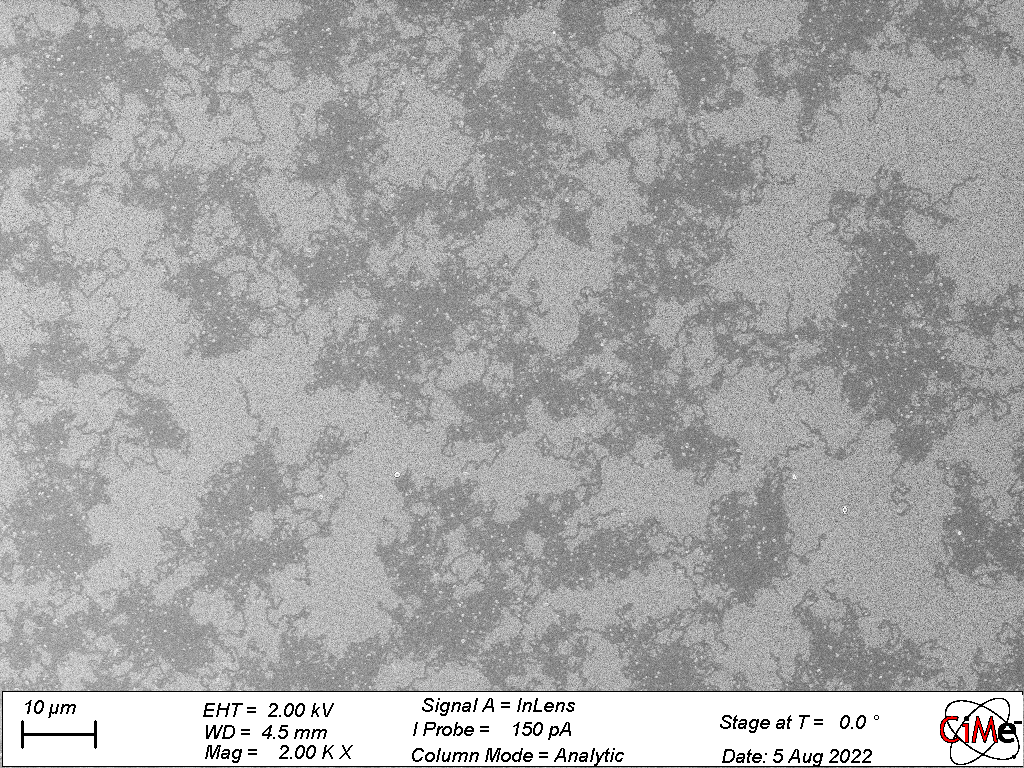}\hfill
			\includegraphics[width=0.49\textwidth]{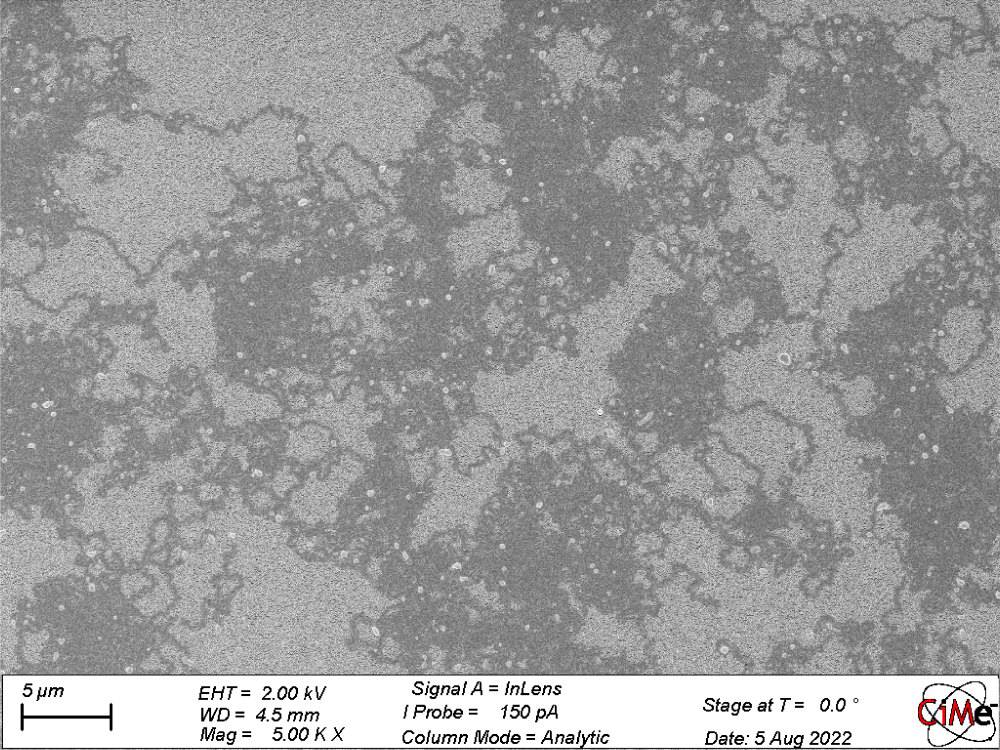}\hfill
			\\[\smallskipamount]
			\includegraphics[width=0.49\textwidth]{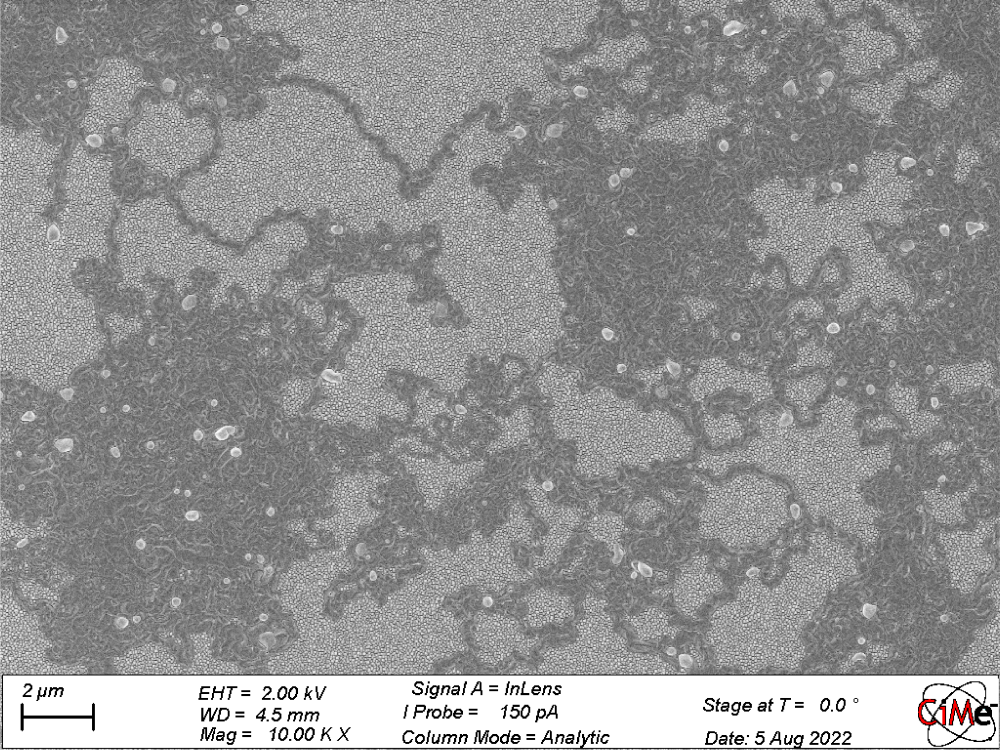}\hfill
			\includegraphics[width=0.49\textwidth]{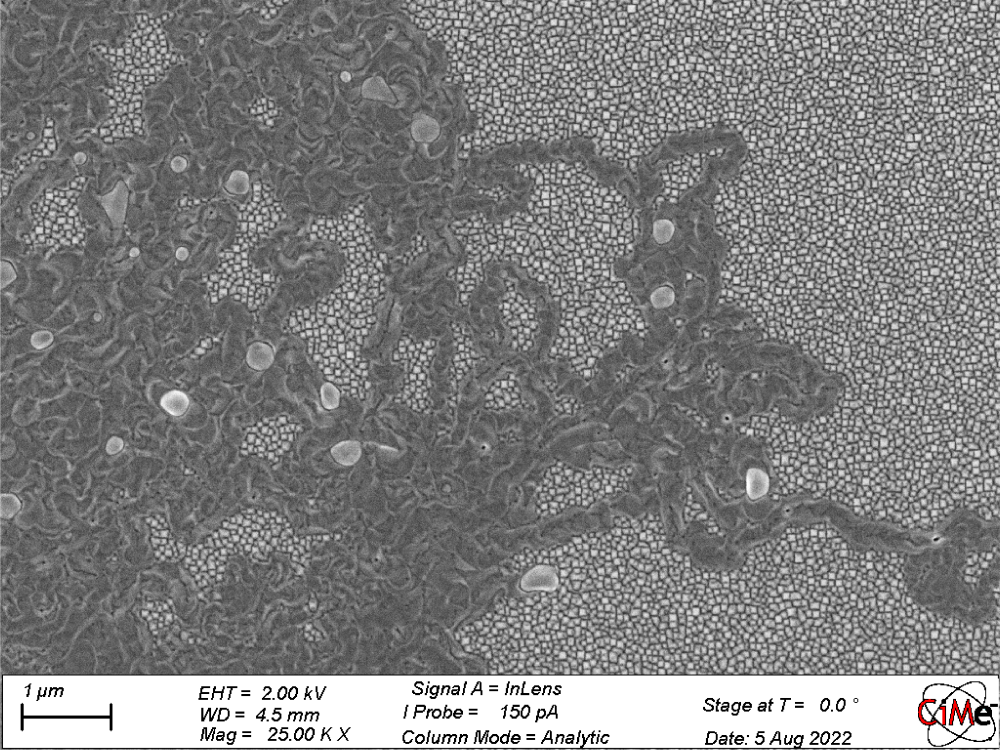}\hfill
			\\[\smallskipamount]
			\includegraphics[width=0.49\textwidth]{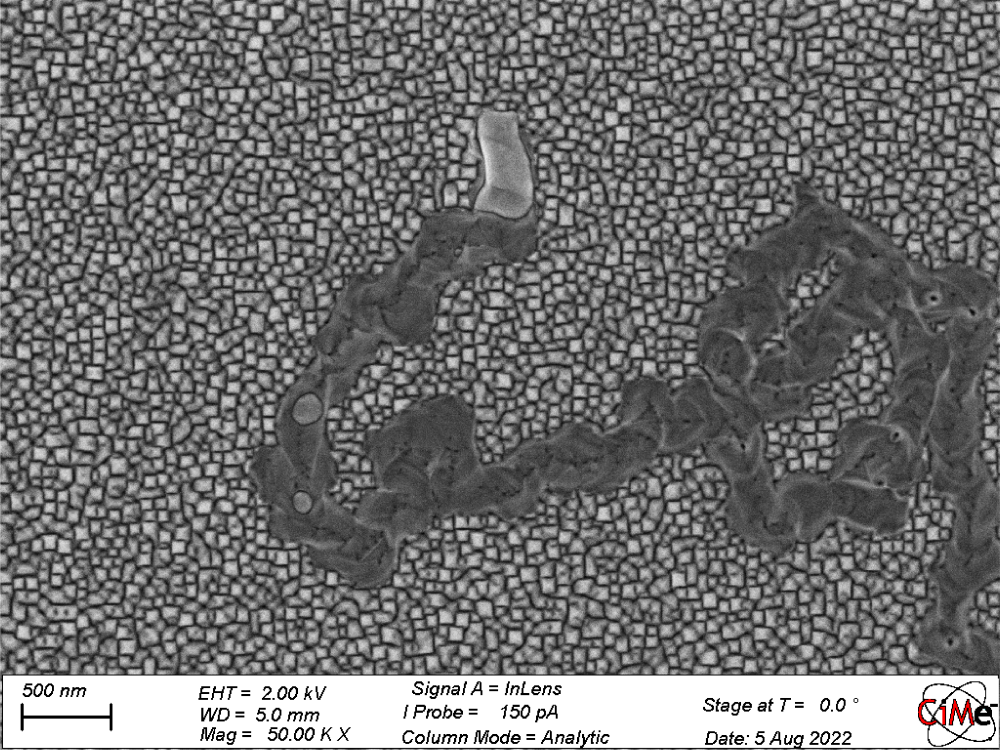}
		\end{figure}

		\begin{figure}[h!]
			\caption{\label{figSI_STEM_EDX}STEM EDX measurements of sample~F with indications of the regions probed to calculate the STEM EDX compositions reported in Fig.~2. A full table of compositions and the measured errors is also reported here. EDX pixel resolution was 10nm, and compositions were corrected for absorption in the TEM lamella thickness using Bote-Salvat ionization cross-sections.}
			\includegraphics[width=0.49\textwidth]{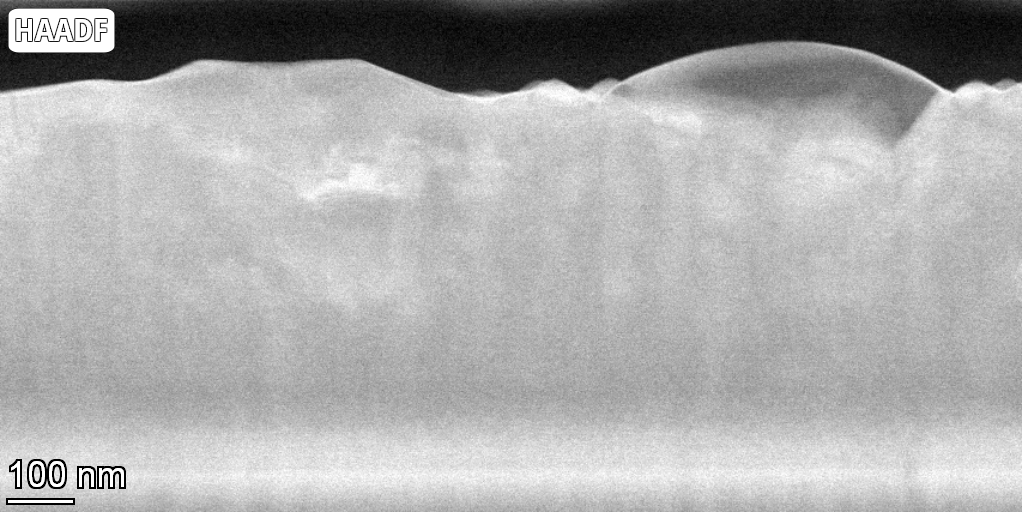}\hfill
			\includegraphics[width=0.49\textwidth]{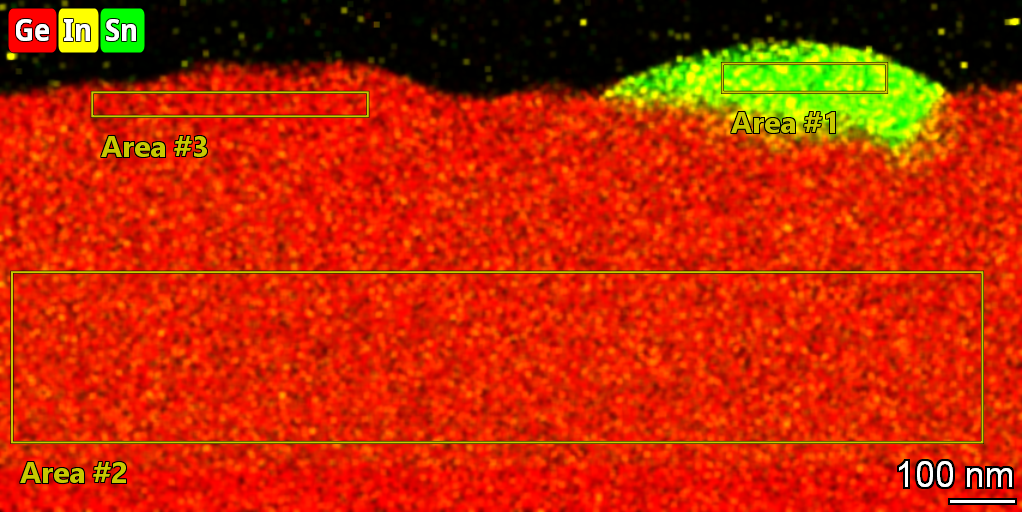}\hfill
			\\[\smallskipamount]
			\includegraphics[width=0.49\textwidth]{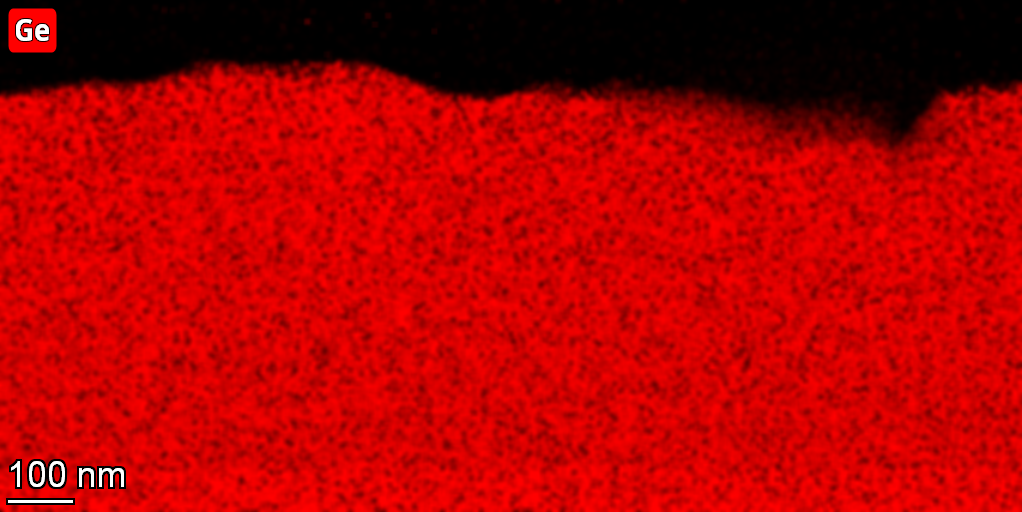}\hfill
			\includegraphics[width=0.49\textwidth]{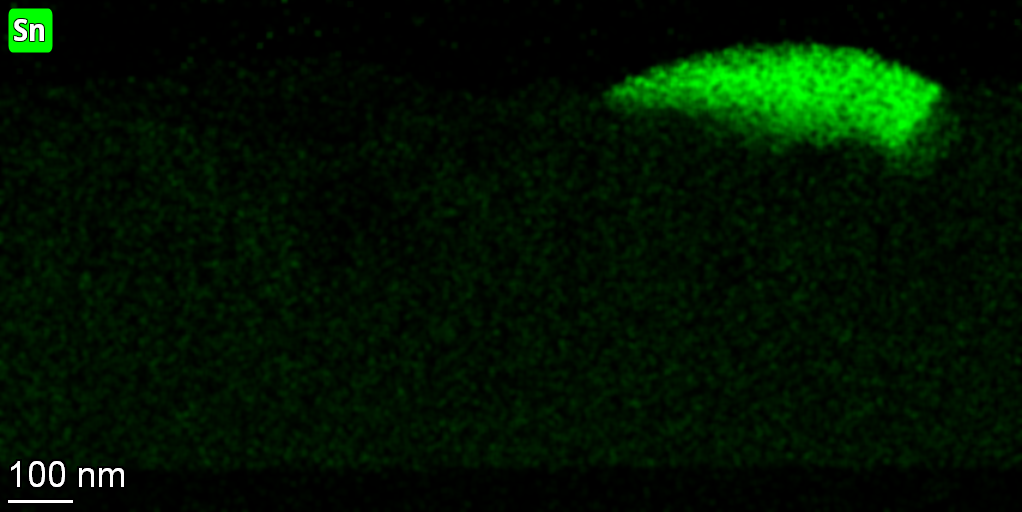}\hfill
			\\[\smallskipamount]
			\includegraphics[width=0.49\textwidth]{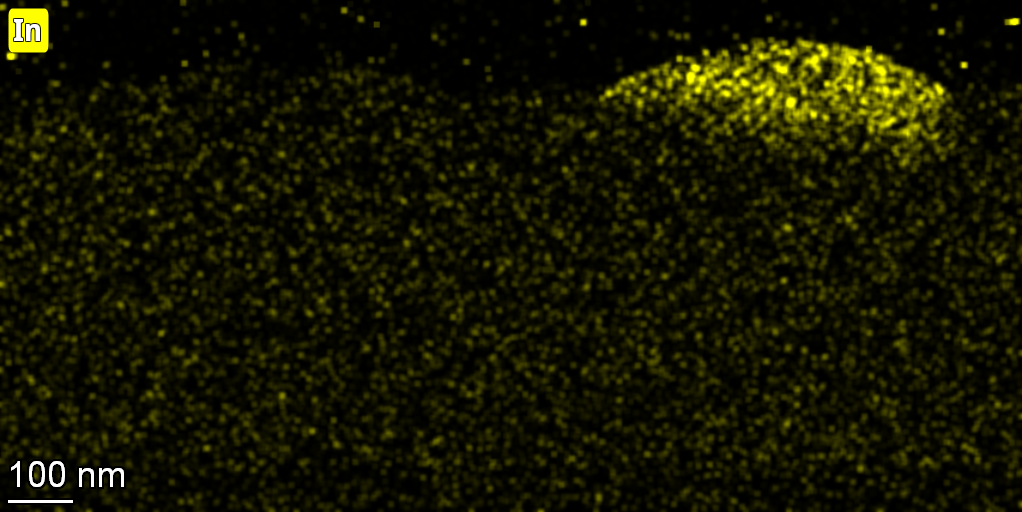}\hfill
			\includegraphics[width=0.49\textwidth]{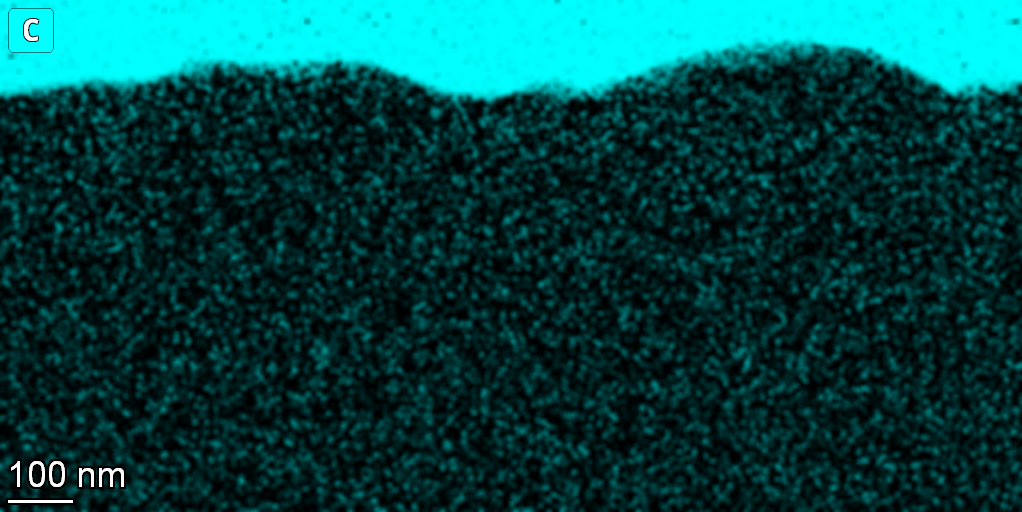}\hfill
			\\[\smallskipamount]
			\includegraphics[width=0.7\textwidth]{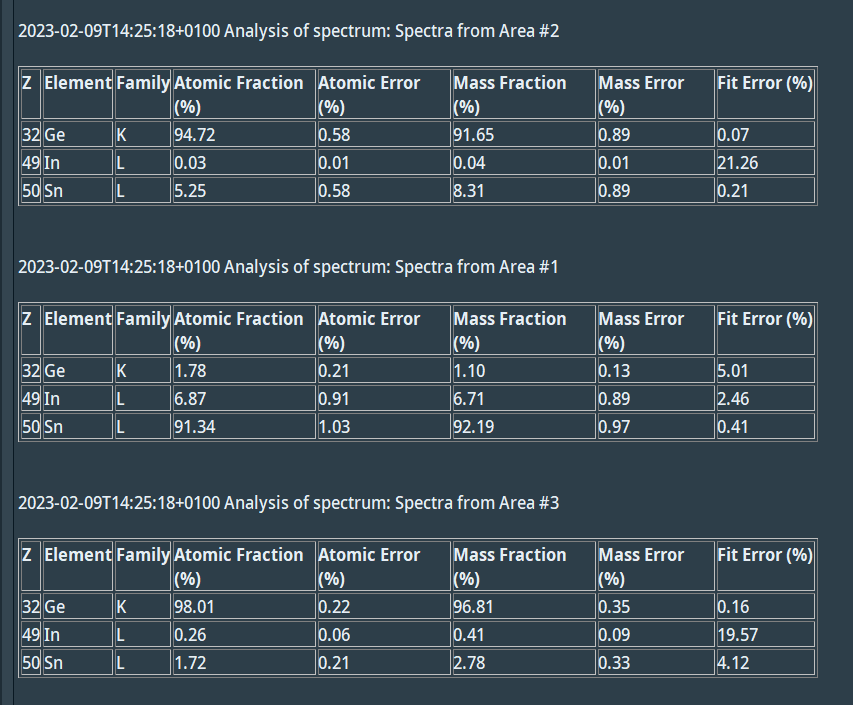}\hfill
		\end{figure}
	
	\begin{figure*}
		\caption{\label{fig_185Cgrowth_SI}
			(a) TEM bright field (BF) image showing a cross-section of Sample~I. (b) Magnified TEM BF, with the positions (yellow circles) of the selected area aperture used for TEM diffraction patterns in Fig.~3c-d. (c) High-resolution TEM image, showing several stacking faults, with fast Fourier transforms taken for two different regions and reported in (d-e). In region \#2, stacking faults are arranged periodically, giving rise to new diffraction spots observed in (e) and in Fig.~3d. On the other hand, in region \#1, stacking faults are not arranged periodically, and thus give rise to streaks rather than additional diffraction spots, seen in both (d) and Fig.~3c.
			(f) TEM dark field (DF) image of polycrystalline regions, and (g) the relative TEM diffractogram showing polycrystalline diffraction spots (yellow arrows).
		}
		\includegraphics[width=0.85\textwidth]{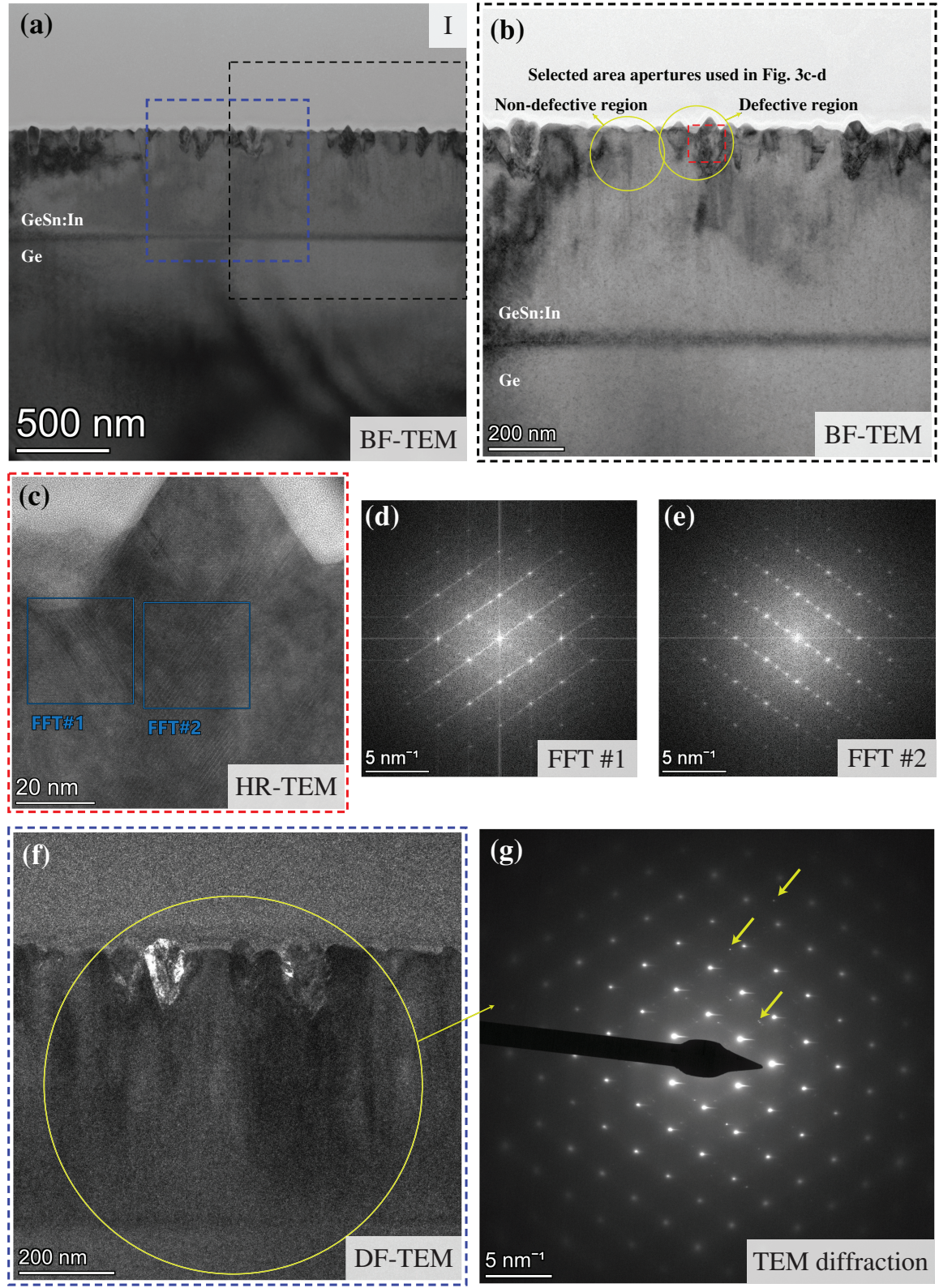}
	\end{figure*} 

	\begin{figure*}
		\caption{\label{SI_DropletsSEMEDX}
		SEM top-view figures of segregation droplets on sample~F, with SEM EDX measurements of droplet composition.
		The ratio in atomic density between Sn and In ($n_{In}/n_{Sn}$) is calculated for each measurement and report in the table. The measured values of $n_{In}/n_{Sn}$ measured here match well that of Fig.~2.
		On a side note, contrary to STEM EDX measurements in Fig.~2, the measured Ge content in these SEM EDX analysis is affected by the signal from the GeSn:In film below the droplet, yielding an apparent higher concentration of Ge in the segregation droplets.
		}
		\includegraphics[width=\textwidth]{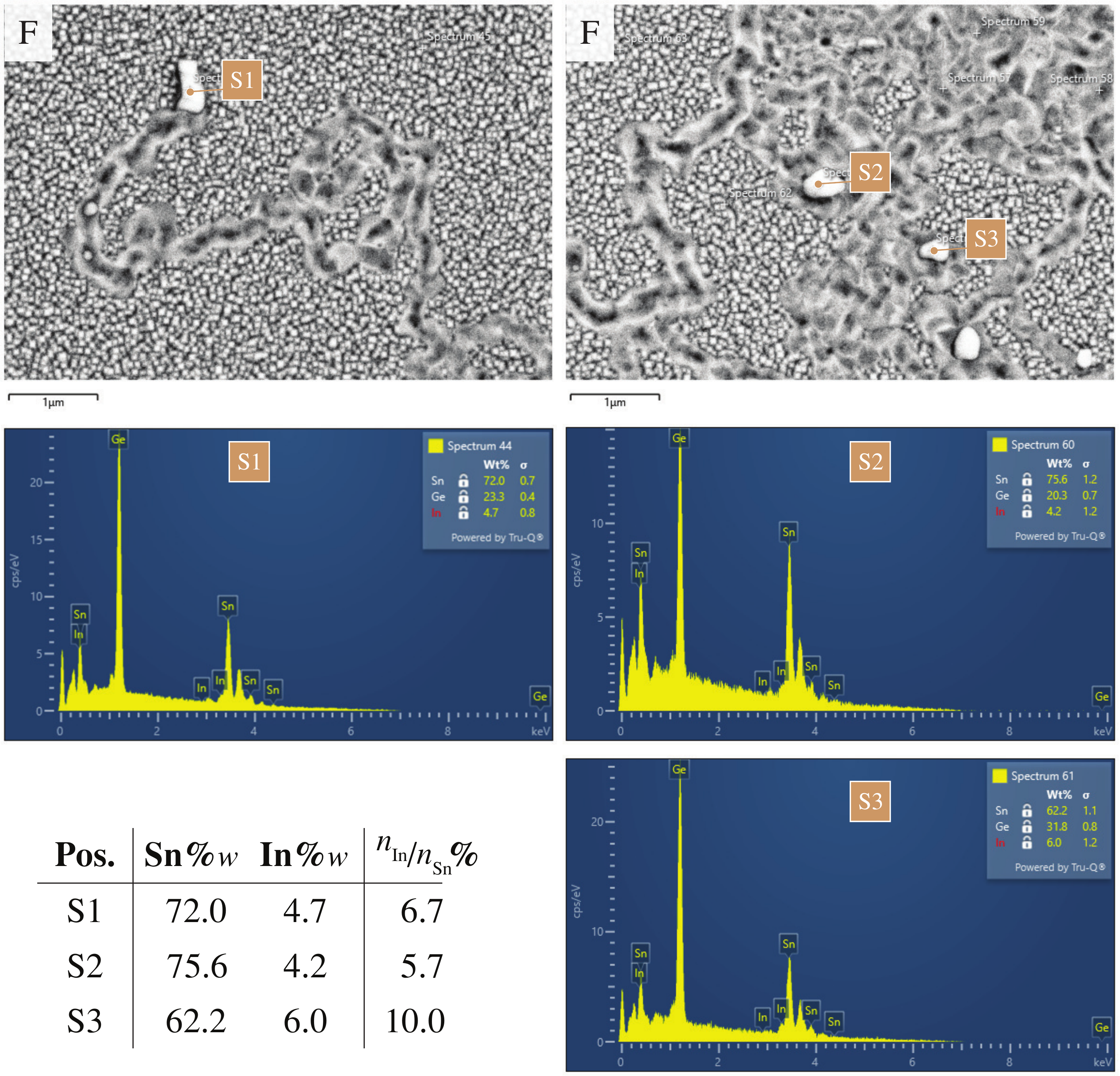}
	\end{figure*} 
	
	%\pagebreak
	%\section*{References}
	%\bibliography{GeSnIn_SI}